\documentclass{aa}  

\usepackage[T1]{fontenc}
\usepackage{xcolor}

\usepackage{natbib}
\usepackage{graphicx} 
\usepackage{amsmath} 
\usepackage{caption}
\usepackage{subfig}
\usepackage{multirow}
\usepackage{comment}
\usepackage{rotating}
\usepackage{ctable}
\usepackage{txfonts}
\usepackage{hyperref}
\begin{document}

   \title{Revealing faint compact radio jets at redshifts above 5 with very long baseline interferometry}


   \author{
   M. Krezinger\inst{1}\fnmsep\inst{2}\fnmsep\inst{3}\fnmsep\thanks{E-mail: krezinger.mate@csfk.org}, 
          G. Baldini\inst{4},
          M. Giroletti\inst{5},
          T. Sbarrato\inst{6},
          G. Ghisellini\inst{6},
          G. Giovannini\inst{4}\fnmsep\inst{5},
          T. An\inst{7}, \\
          K. \'E. Gab\'anyi\inst{1}\fnmsep\inst{9}\fnmsep\inst{2}\fnmsep\inst{3},
          S. Frey\inst{2}\fnmsep\inst{3}\fnmsep\inst{8}
    }

   \institute{
   Department of Astronomy, Institute of Physics and Astronomy, ELTE E\"otv\"os Lor\'and University, P\'azm\'any P\'eter s\'et\'any 1/A, H-1117 Budapest, Hungary  
   \and
    Konkoly Observatory, HUN-REN Research Centre for Astronomy and Earth Sciences, Konkoly Thege Mikl\'{o}s \'{u}t 15-17, H-1121 Budapest, Hungary
    \and
    CSFK, MTA Centre of Excellence, Konkoly Thege Mikl\'os \'ut 15-17, H-1121 Budapest, Hungary
    \and
    Dipartimento di Fisica e Astronomia, Università di Bologna, Via P. Gobetti 93/2, 40129 Bologna, Italy
    \and
    INAF Istituto di Radioastronomia, via Gobetti 101, 40129, Bologna, Italy
    \and
    INAF Osservatorio Astronomico di Brera, Via E. Bianchi 46, 23807, Merate, Italy
    \and
    Shanghai Astronomical Observatory, Key Laboratory of Radio Astronomy and Technology, Chinese Academy of Sciences, 80 Nandan Road, Shanghai 200030, China
    \and
    Institute of Physics and Astronomy, ELTE E\"otv\"os Lor\'and University, P\'azm\'any P\'eter s\'et\'any 1/A, H-1117 Budapest, Hungary
    \and
    HUN-REN--ELTE Extragalactic Astrophysics Research Group, E\"otv\"os Lor\'and University, P\'azm\'any P\'eter s\'et\'any 1/A, H-1117 Budapest, Hungary
    }

   \date{Received 7 June 2024 / Accepted 21 July 2024}

\abstract 
   {Over the past two decades, our knowledge of the high-redshift $(z>5)$ radio quasars has expanded, thanks to dedicated high-resolution very long baseline interferometry (VLBI) observations. Distant quasars provide unique information about the formation and evolution of the first galaxies and supermassive black holes in the Universe. Powerful relativistic jets are likely to have played an essential role in these processes. However, the sample of VLBI-observed radio quasars is still too small to allow meaningful statistical conclusions.}
   {We extend the list of the VLBI observed radio quasars to investigate how the source structure and physical parameters are related to radio loudness.}
   {We assembled a sample of ten faint radio quasars located at $5 < z < 6$ with their radio-loudness indices spanning between $0.9-76$. We observed the selected targets with the European VLBI Network (EVN) at 1.7~GHz. The milliarcsecond-scale resolution of VLBI at this frequency allowed us to probe the compact innermost parts of radio-emitting relativistic jets. In addition to the single-band VLBI observations, we collected single-dish and low-resolution radio interferometric data to investigate the spectral properties and variability of our sources.}
   {The detection rate of this high-redshift, low-flux-density sample is $90\%$, with only one target (J0306$+$1853) remaining undetected. The other nine sources appear core-dominated and show a single, faint and compact radio core on this angular scale. The derived radio powers are typical of Fanaroff-Riley II radio galaxies and quasars. By extending our sample with other VLBI-detected $z > 5$ sources from the literature, we found that the core brightness temperatures and monochromatic radio powers tend to increase with radio loudness.}
   {}



\titlerunning{Faint radio jets at $z>5$ with VLBI}
\authorrunning{M. Krezinger et al.}

   \maketitle
      
%
\section{Introduction} \label{intorduction}
There are currently several hundred quasars known at redshift $z > 5$ and their number is increasing \citep[e.g.][]{1999AJ....118....1F, 2016ApJS..227...11B, 2016ApJ...833..222J, 2019ApJ...884...30W,2022MNRAS.511..572O,2024AJ....168...58D,2023AJ....165..191Y}. Their studies provide essential information about how galaxies formed and evolved in the early Universe. Quasars can be divided into two populations based on their radio-loudness parameter, i.e. the radio-to-optical power ratio usually expressed as $R_{\mathrm{4400\,\AA}} = L_{\mathrm{5\,GHz}}/L_{\mathrm{4400\,\AA}}$ \citep{1989AJ.....98.1195K, 2016ApJ...831..168K} or $R_{\mathrm{2500\,\AA}} = L_{\mathrm{5\,GHz}}/L_{\mathrm{2500\,\AA}}$ \citep{1980ApJ...238..435S}. 
Radio-loud quasars (RLQs) are defined with $R > 10$ and they represent only $\sim 10\%$ of the known quasars \citep[e.g.][]{1984RvMP...56..255B,2002AJ....124.2364I}, also at high redshifts \citep{2015ApJ...804..118B}.
Their radio emission originates mainly from powerful relativistic plasma jets \citep[e.g.][]{2015MNRAS.452.1263P,2019ARA&A..57..467B}. On the other hand, the nature of the the emission of radio-quiet quasars (RQQs, $R \leq 10$) is less obvious. RQQs make up the majority of the active galactic nuclei (AGN) population at every redshift, therefore their study is crucial for understanding the nature of AGN in general. They are thought not to host high-power jets. Their radio emission generally originates from star-forming activity in their host galaxy, accretion disk winds, coronal emission, or shock regions due to the interaction between the outflows and interstellar clouds \citep{2019NatAs...3..387P}. These mechanisms operate at different spatial scales and radio frequencies. Recently, \citet{2021AA...655A..95S} revealed that at high redshifts, some massive ($M>10^{10}\,\mathrm{M_{\odot}}$) RQQs host powerful relativistic jets, which seems to be at odds with the above model. 

The distant AGN are known to host supermassive black holes (SMBHs) with $M_{\mathrm{BH}} > 3 \times 10^{8}\,\mathrm{M_{\odot}}$, with the majority being more massive than $10^{9}\,\mathrm{M_{\odot}}$ \citep[e.g.][]{2013ApJ...773...44W,2019ApJ...873...35S,2021Galax...9...23S}. Detecting preferentially these objects is clearly a selection bias, as they are the most powerful sources of radiation, easily observable even from vast distances corresponding to high redshifts. The presence of $M_{\mathrm{BH}} \gtrsim 10^{8-9}\,\mathrm{M_{\odot}}$ black holes raises important questions about the formation and evolution of the first SMBHs in the Universe. \citet{2012Sci...337..544V} and \citet{2020ARA&A..58...27I} reviewed the possible mechanisms to build up SMBHs in excess of $10^{9}\,\mathrm{M_{\odot}}$ in less than 1~Gyr. It can only be achieved via fast accretion in a super-Eddington regime, or assuming black hole seeds of $\sim10^{4}\,\mathrm{M_{\odot}}$ as a starting point not to violate the Eddington limit. Jets might play an important role in assembling such massive early black holes. This can be explained by a simple model \citep[e.g.][]{2013MNRAS.432.2818G,2021Galax...9...23S,2022AA...663A.147S} where the relativistic jets carry away some of the released gravitational energy, allow for a faster accretion at the same observed luminosity. The distribution of jets in the early Universe, whether they are connected to RLQs or RQQs, is important to understand their link to the formation of the first SMBHs. 

Another intriguing observational result regarding the high-redshift AGN, inferred by the presence of blazars (i.e. sources with their jets aligned close to the line of sight, $\theta \lesssim 10\degr$), is that the fraction of the most massive ($>10^{9}\,\mathrm{M_{\odot}}$) black holes hosted in radio-loud (jetted) AGN is larger than those found in RQQs at $z>4$ compared to more recent cosmological epochs \citep{2013MNRAS.432.2818G,2015MNRAS.446.2483S,2021Galax...9...23S}. It appears that high-redshift AGN with the most massive black holes prefer to have relativistic jets. There are a few scenarios that can explain the problem of missing RQQs. \citet{2011MNRAS.416..216V} proposed the idea that the difference might be due to internal and external absorption mechanisms in AGN, or lower bulk Lorentz factors in jets in the early Universe. It is also possible that the current survey sensitivity causes bias. At high redshift, \citet{2015MNRAS.452.3457G} found the radio lobes dimmed due to the interaction of their electrons with photons of the cosmic microwave background. The early galaxies may also be obscured by a dense dust bubble surrounding the central region of the AGN \citep{2016MNRAS.461L..21G} and only the most powerful jets are able to penetrate this bubble, sweeping away most of the material along their path. The absence of direct link between radio-loudness and jet presence raises the question whether there is a missing population of relativistic jets hosted in RQQs. Or we are simply misclassifying a fraction of blazar candidates based on X-ray and low-resolution radio observations \citep[e.g.][]{2017MNRAS.467..950C,2016MNRAS.463.3260C,2022ApJS..260...49K}.

To reliably observe jets in RQQs, parsec-scale resolution is required, because the radio emission on this scale is closely associated with the central engine. On the other hand, low-resolution observations (arcsecond-scale) trace the total radio flux density, a composite emission of the different mechanisms, which cannot be directly linked to the optical, X-ray, or infrared emissions coming from the AGN. However, low-resolution interferometric surveys are the starting point for any follow-up radio jet study. The Faint Images of the Radio Sky at Twenty-Centimeters \citep[FIRST,][]{1997ApJ...475..479W} survey catalogue at 1.4-GHz and the recent, still ongoing multi-epoch Very Large Array Sky Survey \citep[VLASS,][]{2020PASP..132c5001L,2020RNAAS...4..175G,2021ApJS..255...30G} provide essential spectral and structural information in help to select targets for high-resolution observations. The technique of very long baseline interferometry (VLBI) can achieve parsec-scale linear resolution at cm wavelengths and is capable to distinguish between compact emission with high brightness temperature and a more extended radio feature. Only a few dozens of RQQs have been observed with VLBI \citep{1998MNRAS.299..165B,2004A&A...417..925M,2005ApJ...621..123U,2009ApJ...706L.260G,2013MNRAS.432.1138P,2016A&A...589L...2H,2020Symm...12..527K,2023MNRAS.518...39W}, and very few that can be found at high redshift \citep[i.e. $z > 4$,][]{2009MNRAS.398..176K,2021AA...655A..95S}.

VLBI can provide useful astrometric information as well. Phase-referenced VLBI observations  \citep{1995ASPC...82..327B}, where nearby calibrator sources provide relative astrometric positions for the targeted radio-emitting features, can be accurate to milliarcsecond (mas) level. This accuracy is comparable to the precise optical astrometric data from the recent 3rd Data Release (DR3) of the \textit{Gaia} mission \citep{2016A&A...595A...1G, 2023A&A...674A...1G}. The optical position, if available, in addition to the VLBI radio position, adds further relevant astrophysical and morphological information about the source. The optical position marks the location of the AGN accretion disk or, in some cases, its blend with the optical synchrotron emission of the innermost sub-milliarcsecond scale jet \citep{2019ApJ...871..143P}. Unlike the optical, the VLBI intensity peak pinpoints the brightest and most compact emission feature, the self-absorbed base of the jet (i.e. the core), or sometimes a shock front (i.e. a hotspot in a lobe) in an extended radio source \citep{2017A&A...598L...1K}. 

In this paper, we present VLBI observations of ten $z > 5$ radio quasars with different radio-loudness indices. This way, we expect to cover a variety of radio-emitting AGN in order to understand whether the radio-loudness cut is meaningful in terms of the presence of relativistic jets at such high redshifts. The experiment was carried out with the European VLBI Network (EVN) combined with e-MERLIN (Enhanced Multi-Element Radio-Linked Interferometer Network) antennas at the frequency of 1.7~GHz in 2022. The sensitivity and high angular resolution of the EVN allow us to detect compact core--jet features. With the inclusion of e-MERLIN stations, it is also possible to detect extended ($\sim 0\farcs1$) emission if present. The goal was to investigate the nature of these sources using high-resolution VLBI imaging data, together with radio spectral information found in the literature, and the recent \textit{Gaia} DR3 optical positions where available. By involving literature data about other known $z>5$ quasars as well, we investigate how the milliarcsecond-scale properties are related to the radio loudness of the sources.

In Section~\ref{sample}, we introduce the sample of ten objects targeted with the EVN. In Section~\ref{obs&reduc}, the details of the observations and data reduction are provided. Our result are given in Section~\ref{results} and discussed in Section~\ref{discussion}. The findings of this study are summarised in Section~\ref{conclusion}. Throughout this paper, we assume a standard flat $\Lambda$ Cold Dark Matter cosmological model with $\Omega_{\mathrm m} = 0.3$, $\Omega_{\mathrm \Lambda} = 0.7$, and $H_0 = 70$~km\,s$^{-1}$\,Mpc$^{-1}$, and used these parameters in the cosmology calculator of \citet{2006PASP..118.1711W}. In this model, the typical linear scale is $\sim6$~pc\,mas$^{-1}$ and the luminosity distance of our sources is $\sim50$~Gpc.

\section{Target sample} \label{sample}

With this study, we aim to extend the list of the VLBI-observed high-redshift radio-quiet quasars. To this end, we searched for faint radio emitters among the sources in the catalogue of the known $z > 5$ quasars compiled by \citet{2020MNRAS.494..789R}. From this catalogue, the nine sources with the lowest radio power were selected. These have VLASS detection but have not been observed with VLBI yet. Their 2.7-GHz VLASS flux densities are ranging between $1\,\mathrm{mJy} < S_{\mathrm{2.7\,GHz}} < 11\,\mathrm{mJy}$. Six of them were also detected in the 1.4-GHz FIRST survey. The 10th target, SDSS J0306$+$1853, having one of the most massive known SMBHs among quasars at $z > 5$ with 1.1$\times10^{10}$ M$_{\odot}$  \citep{2015ApJ...807L...9W}, is also included in the sample. It was previously investigated by \citet{2021AA...655A..95S} and found to be unresolved on arcsecond scales with the \textit{Karl G. Jansky} Very Large Array (JVLA), suggesting the presence of a compact radio jet. Its flux densities were 0.25~mJy and 0.09~mJy at 1.5 and 5~GHz frequencies, respectively \citep{2021AA...655A..95S}. There were no other radio observations of this quasar.

Table~\ref{tab:sample} contains the parameters of the selected radio sources. Since all but one of them have VLASS detection, for the sake of simplicity and uniformity, we calculated the radio loudness index using the 2.7-GHz flux density as $R = L_{\mathrm{2.7\,GHz}} / L_{\mathrm{4400\,A}}$. In the case of J0306$+$1853 which is not detected in VLASS, we used the flux densities and spectral index derived by \citet{2021AA...655A..95S} to estimate the 2.7-GHz radio power. The 10 sources in our sample have radio loudness in the range $0.9 \leq R \leq 76$.

\begin{table*}[!ht]
    \caption{Target sources and information on their respective phase-reference calibrators used in the EVN observations.}       
    \label{tab:sample}     
    \centering                          
    \begin{tabular}{ccccccc|cc}       
        \hline\hline               
    Source ID & $z$ & $D_\mathrm{L}$ & Linear scale & $S_{\mathrm{2.7}}$ & $S_{\mathrm{1.4}}$ & $R$ & Phase & Separation \\  
              &  & [Mpc]  & [pc\,mas$^{-1}$]  & [mJy] & [mJy] & [$L_{\mathrm{2.7\,GHz}}/L_{\mathrm{4400\,A}}$] &  calibrator & [\degr] \\
        \hline                      
    J0306$+$1853 & 5.363$^1$ & 50596 & 6.06 & 0.15 (0.01)$^*$ & ... & 0.9 & J0259+1925 & 1.79 \\
    J0616$-$1338 & 5.580$^2$ & 53030 & 5.94 & 1.3 (0.2) & ... & 9.3  & J0618$-$1418 & 0.80  \\
    J0741$+$2520 & 5.194$^3$ & 48767 & 6.16 & 4.2 (0.2) & 2.9 (0.1) & 6.5 & J0746+2549 & 1.12 \\
    J0747$+$1153 & 5.260$^4$ & 49493 & 6.12 & 1.5 (0.2) & ... & 8.6  & J0749+1057 & 1.02 \\
    J0901$+$1615 & 5.630$^5$ & 53585 & 5.91 & 2.9 (0.4) & 3.9 (0.2) & 76.1 & J0905+1541 & 1.10 \\
    J1034$+$2033 & 5.010$^6$ & 46748 & 6.28 & 3.9 (0.3) & 3.9 (0.1) & 20.4 & J1036+2203 & 1.59 \\
    J1614$+$4640 & 5.313$^6$ & 50077 & 6.09 & 4.0 (0.2) & 1.7 (0.1) & 20.3 & J1616+4632 &  0.31\\
    J2239$+$0030 & 5.090$^6$ & 47624 & 6.23 & 1.0 (0.2) & 1.4 (0.1) & 29.9 & J2239+0128 & 0.98 \\
    J2245$+$0024 & 5.170$^6$ & 48393 & 6.18 & 1.4 (0.2) & 0.9 (0.1) & 58.0 & J2247+0000 &  0.66 \\
    J2344$+$1653 & 5.000$^6$ & 46638 & 6.28 & 10.5 (0.2) & ... & 24.3 & J2347+1709 & 0.85 \\
        \hline                               
\end{tabular}
\newline
Notes. Col.~1 -- source name derived from J2000 equatorial coordinates; Col.~2 -- spectroscopic redshift and its reference in the upper index (1: \citet{2015ApJ...807L...9W}, 2: \citet{2019ApJ...871..199Y}, 3: \citet{2009AJ....138.1925M}, 4: \citet{2016ApJ...829...33Y}, 5: \citet{2015ApJ...804..118B}, 6: \citet{2016ApJ...819...24W}); Col.~3 -- luminosity distance; Col.~4 -- linear scale at the source redshift;
Col.~5 -- VLASS 2.7-GHz flux density (uncertainty in parentheses). $^*$ The 2.7-GHz flux density calculated using the spectral index from \citet{2021AA...655A..95S}; Col.~6 -- FIRST 1.4-GHz flux density (uncertainty in parentheses); Col.~7 -- radio-loudness parameter; Col.~8 -- phase-reference calibrator source name;  Col.~9 -- angular separation between the target and calibrator sources.
\end{table*}

\section{Observations and data reduction} \label{obs&reduc}

\subsection{EVN observations}   \label{observations}

\begin{figure}[!ht]
    \centering
    \includegraphics[width=0.40\textwidth]{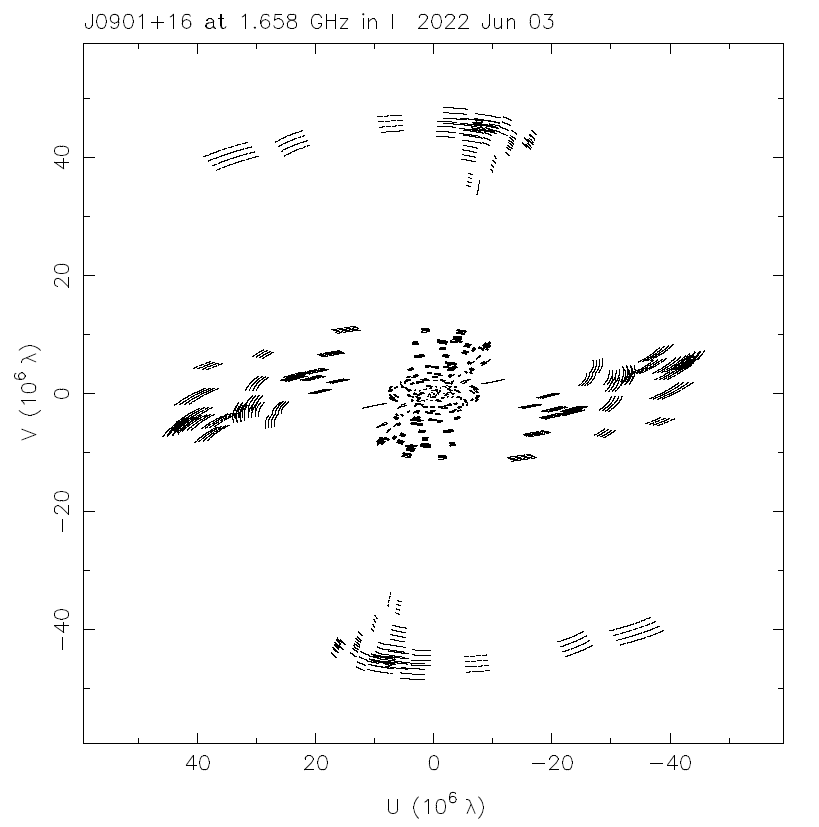}
    \includegraphics[width=0.40\textwidth]{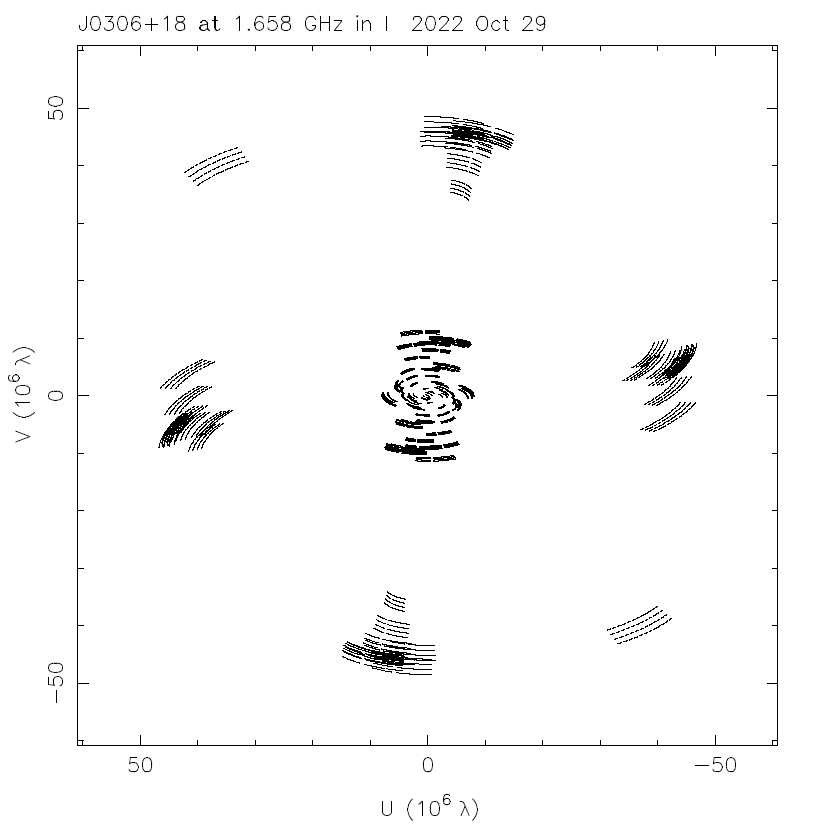}
    \caption{Typical $(u,v)$ coverages for low-declination targets in the EG119A (\textit{top}) and EG119B (\textit{bottom}) project segments.}
    \label{fig:uvpl}
\end{figure}

The selected ten faint radio sources were observed with the EVN at a single frequency band centred around 1.66~GHz. The observations were carried out in two sessions under the project code EG119 (PI: G. Baldini), one with nine targets in 2022 June 3 (EG119A) and another with J0306$+$1853 as the only target in 2022 October 22 (EG119B). In addition to the elements of the EVN, antennas of the e-MERLIN were also included in the VLBI network. Since the targets are faint sources, it was necessary to use the technique of phase referencing \citep{1995ASPC...82..327B}. With phase-referencing, in addition to increasing the coherent integration time on faint sources, we can also relate target positions to the known ICRF \citep[Internatonial Celestial Reference Frame,][]{2020A&A...644A.159C} astrometric positions of the respective calibrator sources. In this observing mode, the radio telescopes were nodding between a nearby bright compact calibrator source and the given target, in repeating cycles. In our case, within the 8.5-min cycle time, 6.5~min was spent on the weak target source, the rest was spent on the calibrator and with antenna slewing. The EG119B run had a shorter, 7-min cycle time, of which 4.6~min was spent on the target source. The total bandwidth of 128~MHz was divided into four 32-MHz wide intermediate frequency channels (IFs), each further divided into 64 spectral channels. The data were recorded with 1024~Mbps data rate, in left and right circular polarizations. Bright compact fringe-finder sources were also scheduled in the experiments, J0646$+$4451, J1642$+$3948, and J2253$+$1608 in EG119A, and 3C\,84 in EG119B. Overall, 15 radio telescopes participated in the experiments: Jodrell Bank Mk2 ($38\,\mathrm{m} \times 25\,\mathrm{m}$ diameter, United Kingdom), Westerbork ($25$~m, The Netherlands), Effelsberg ($100$~m, Germany), Medicina ($32$~m, Italy), Noto ($32$~m, Italy), Onsala ($25$~m, Sweden), Tianma ($65$~m, China), Urumqi ($25$~m, China), Toru\'{n} ($32$~m, Poland), Hartebeesthoek ($26$~m, South Africa) and the e-MERLIN antennas in the United Kingdom: Cambridge ($32$~m), Darnhall, Defford, Knockin, and Pickmere ($25$~m each). Urumqi participated in the EG119A experiment only. The data were processed at the Joint Institute for VLBI European Research Infrastructure Consortium (JIVE, Dwingeloo, The Netherlands) with the SFXC software correlator \citep{2015ExA....39..259K} with 2~s integration time. Typical $(u,v)$ coverages for the two project segments are shown in Fig.~\ref{fig:uvpl}.

\subsection{Data reduction} \label{reduction}

The data were calibrated using the NRAO Astronomical Image Processing System (\textsc{aips}) software package \citep{2003ASSL..285..109G}, following a standard procedure \citep[e.g.][]{1995ASPC...82..227D}. As the first step after loading the raw correlated data, the interferometric visibility amplitudes were calibrated using the antenna gain curves and the system temperatures measured at each telescope. Then, the data were corrected for the dispersive ionospheric delay using total electron content maps derived from global navigation satellite systems data. Phase changes due to the time variation of the source parallactic angle were also corrected for radio telescopes with azimuth--elevation mount. An initial correction of instrumental phases and delays was performed using a 1-min scan on a strong fringe-finder source.
Then global fringe-fitting \citep{1983AJ.....88..688S} was performed on the phase-reference and fringe-finder calibrators. These calibrated visibility data were exported to the \textsc{Difmap} software package \citep{1994BAAS...26..987S}, where we carried out hybrid mapping. The process involved several iterations of the \textsc{clean} deconvolution algorithm \citep{1974A&AS...15..417H} and phase-only self-calibration \citep{1984ARA&A..22...97P}, followed by a few rounds of amplitude and phase self-calibration with decreasing solution intervals. Antenna-based median gain correction factors, if in excess of $\pm5\%$, were applied to the visibility amplitudes in \textsc{aips}. The \textsc{clean} model components of the calibrators produced in \textsc{Difmap} were transferred to \textsc{aips} as inputs for a repeated fringe-fitting of the calibrator data. By taking the phase-reference calibrator source structure into account, we could improve phase solutions. The fringe-fit solutions obtained for the phase-reference calibrators were interpolated to the respective target source data. The calibrated visibility data of the target sources were then exported from \textsc{aips}.

The final target source images were produced with \textsc{Difmap} \citep{1994BAAS...26..987S}. To reduce the image noise, we applied natural weighting, where weights were calculated as the reciprocal of the amplitude errors (\textsc{uvweight $0,-1$}). The dirty images were shifted to make the brightness peak coincide with the origin at (0,0) relative right ascension and declination. Following a standard mapping procedure for weak (mJy-level) targets, only a few rounds of \textsc{clean} iterations were performed without self-calibration. After reaching an insignificant peak-to-noise level (below $\sim 5\sigma$) in the residual images, we performed a final \textsc{clean} iteration involving $1000$ steps with a very small loop gain $(0.01)$. The purpose of this last step is to smooth the noise features in the final images. The resulting images are presented in Fig.~\ref{fig:maps} and their parameters are listed in Table~\ref{tab:imgparams}. The phase-referenced VLBI positions of the sources were obtained from the images using the \textsc{aips} task \textsc{maxfit}.

The e-MERLIN stations participating in the interferometer network provided relatively short baselines and thus allowed us to check whether there is $\sim100$-mas scale extended emission in our sources. In \textsc{Difmap}, we applied a taper to the visibility data to reduce the weight of the longer baselines (\textsc{uvtaper 0.5, 12}). This way we increased the sensitivity to detect possible extended emission. We set the full width at half maximum (FWHM) of the Gaussian taper at the baseline length of 12 million wavelengths~(M$\lambda$).

In addition to producing the \textsc{clean} images, circular Gaussian brightness distribution model components were also fitted to the visibility data \citep{1995ASPC...82..267P} in \textsc{Difmap}. This allowed us to quantitatively characterise component sizes and flux densities (Table~\ref{tab:physparams}). In the absence of self-calibration, the effect of coherence loss \citep{2010A&A...515A..53M} on the peak brightness and flux density values has to be considered. To estimate the level of correction for coherence loss, we imaged and modelled the calibrator sources with and without performing phase self-calibration, and found an average model flux density difference of $25\%$. Therefore, the fitted flux densities and the peak brightness values were multiplied by a factor of $1.25$. This factor can still be considered as a lower limit because it does not take into account the fact that the calibrator data have been fringe fitted, and their visibility phases are already corrected to a better degree than for the targets. Note that our correction factor is consistent with estimates found for other EVN experiments \citep[e.g.][]{2006A&A...445..413M,2019A&A...630L...5G,2022ApJS..260...49K}. 

The uncertainties of the parameters of the fitted Gaussian model components were estimated following the method of \citet{1999ASPC..180..301F}. For the flux densities, an additional $5\%$ error was added in quadrature, to account for the VLBI absolute amplitude calibration uncertainty \citep[e.g.][]{2012ApJS..198....5A,2015MNRAS.446.2921F}.

\section{Results} \label{results}

\begin{figure*}[!ht]
    \centering
    \includegraphics[width=0.3\textwidth]{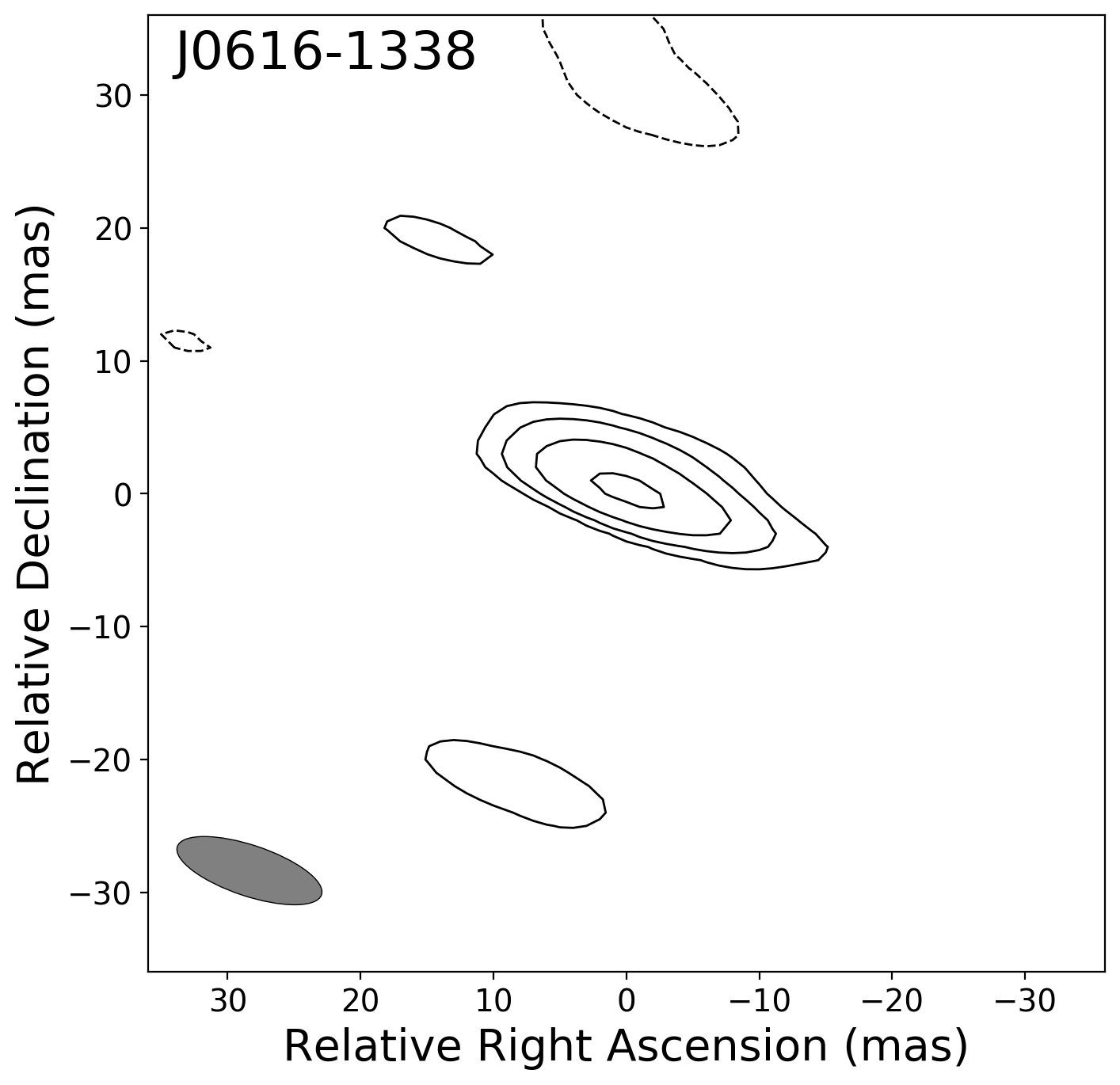}
    \includegraphics[width=0.3\textwidth]{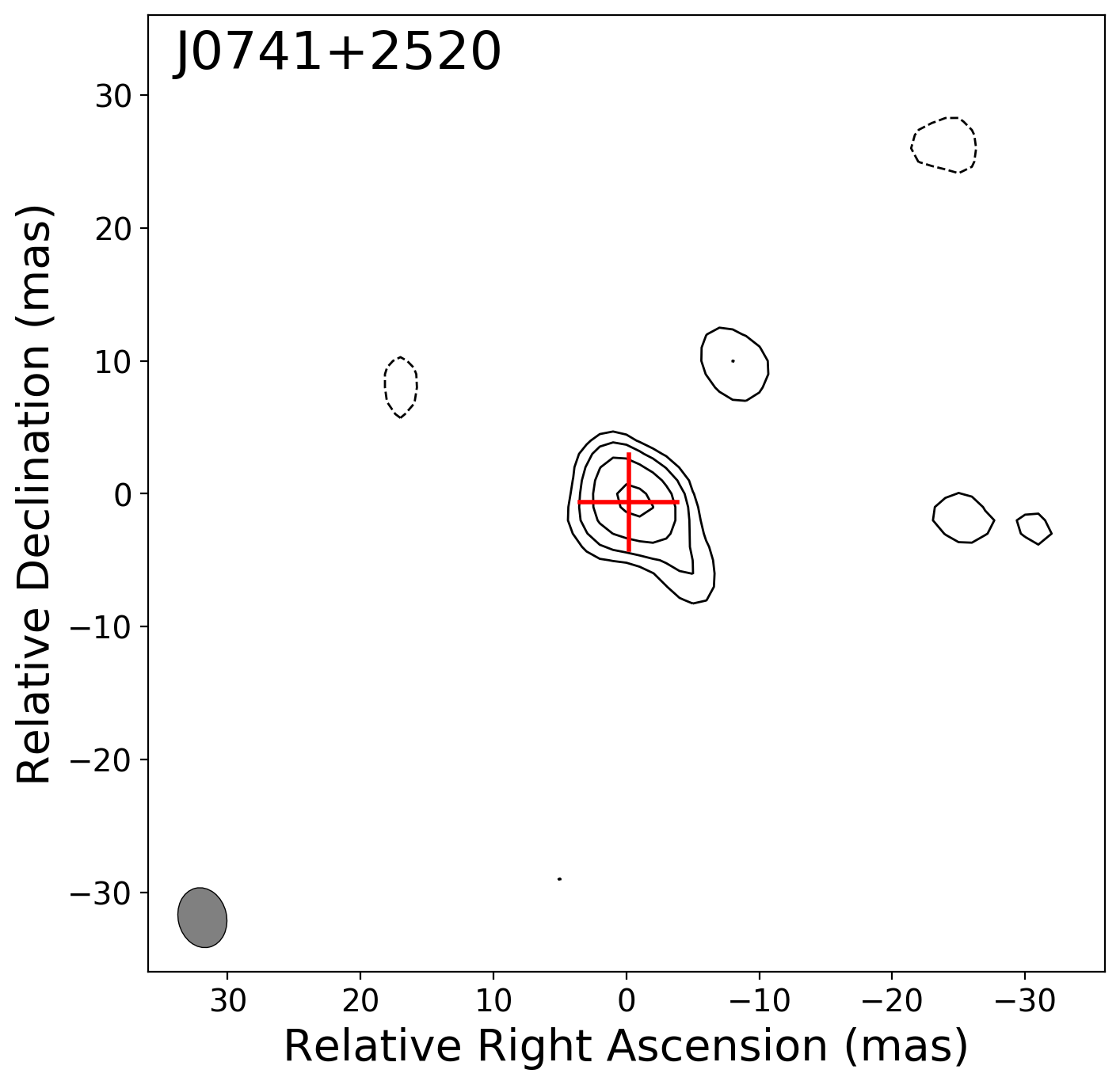}
    \includegraphics[width=0.3\textwidth]{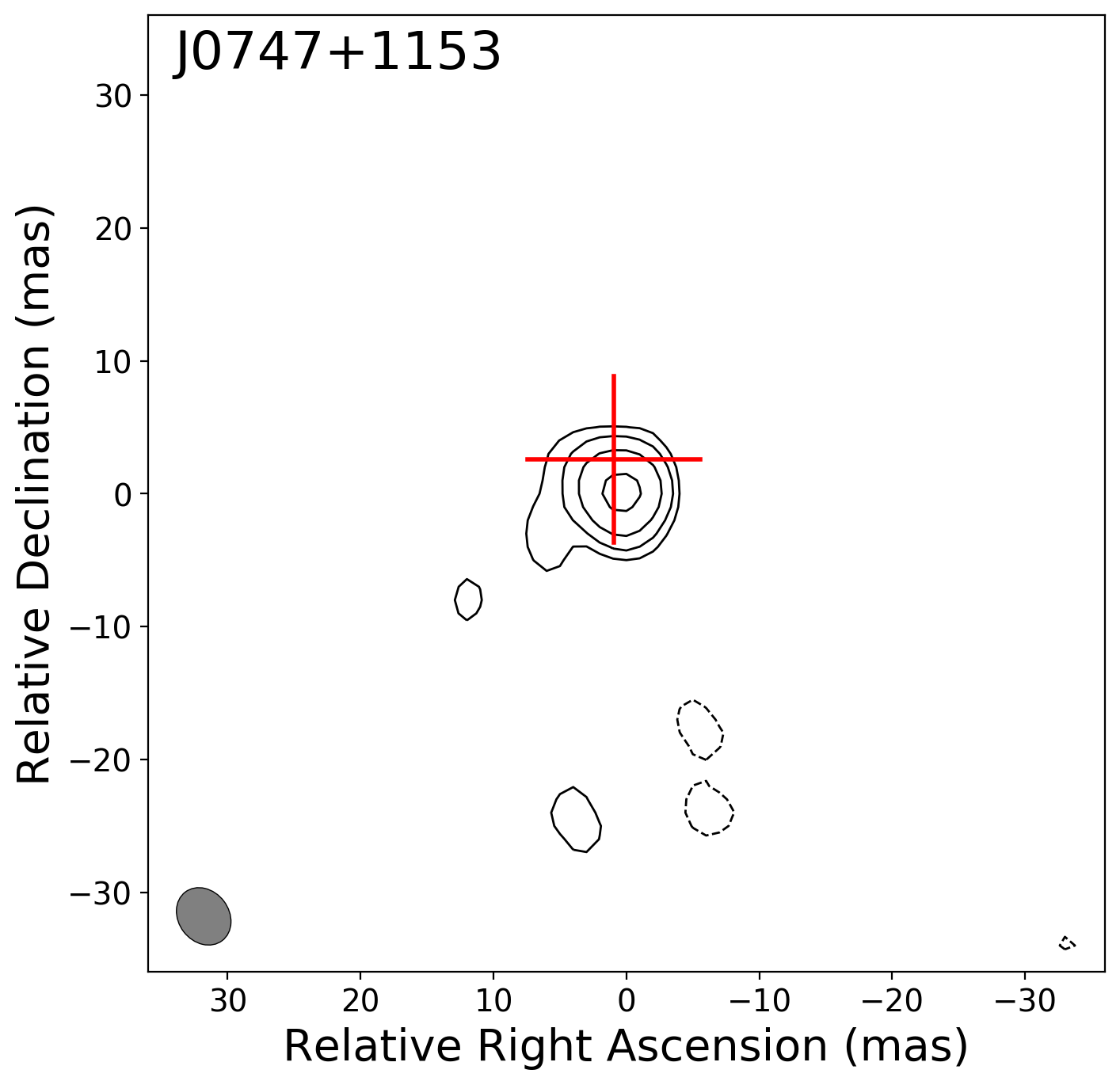}
    \includegraphics[width=0.3\textwidth]{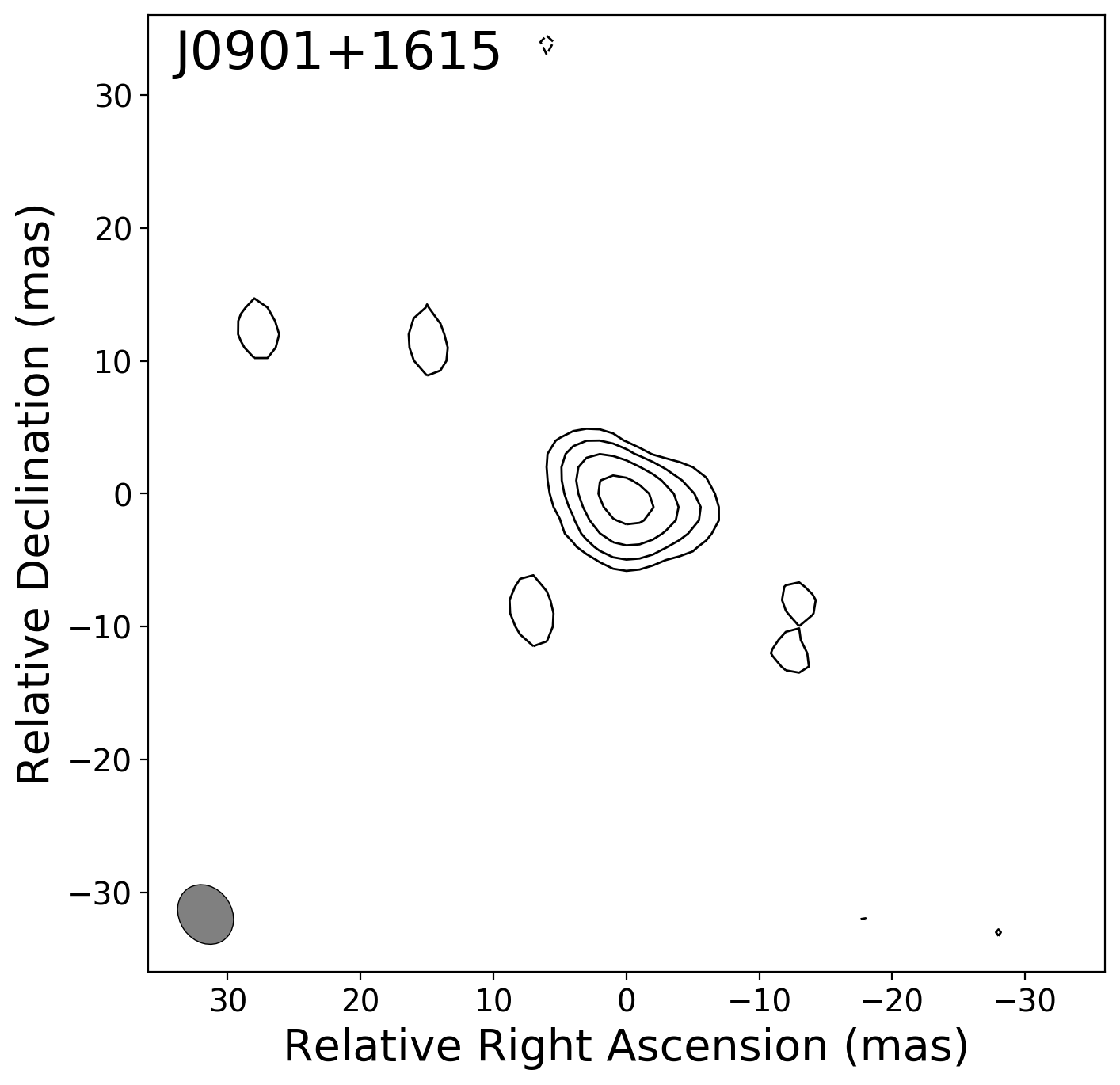}
    \includegraphics[width=0.3\textwidth]{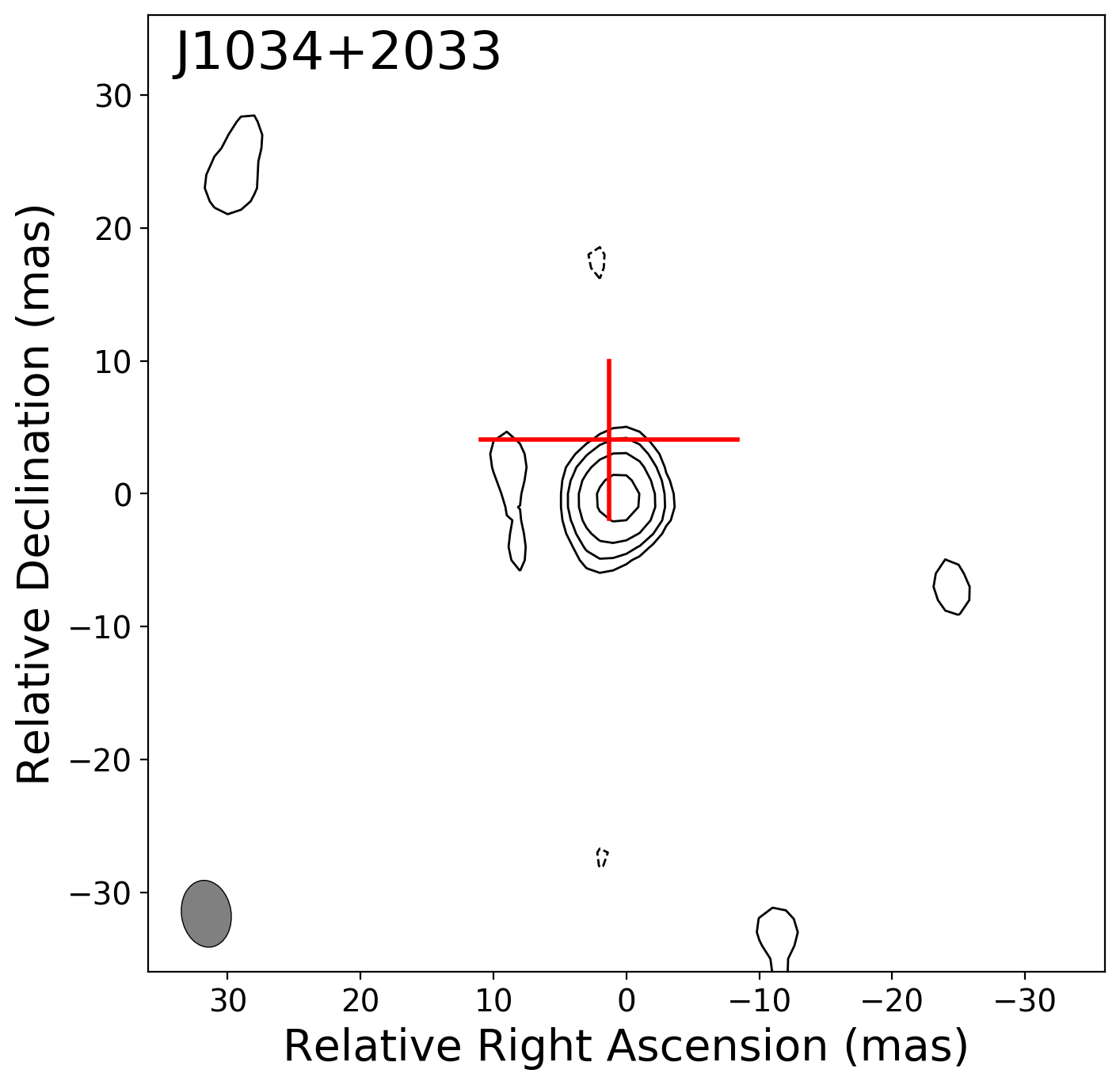}
    \includegraphics[width=0.3\textwidth]{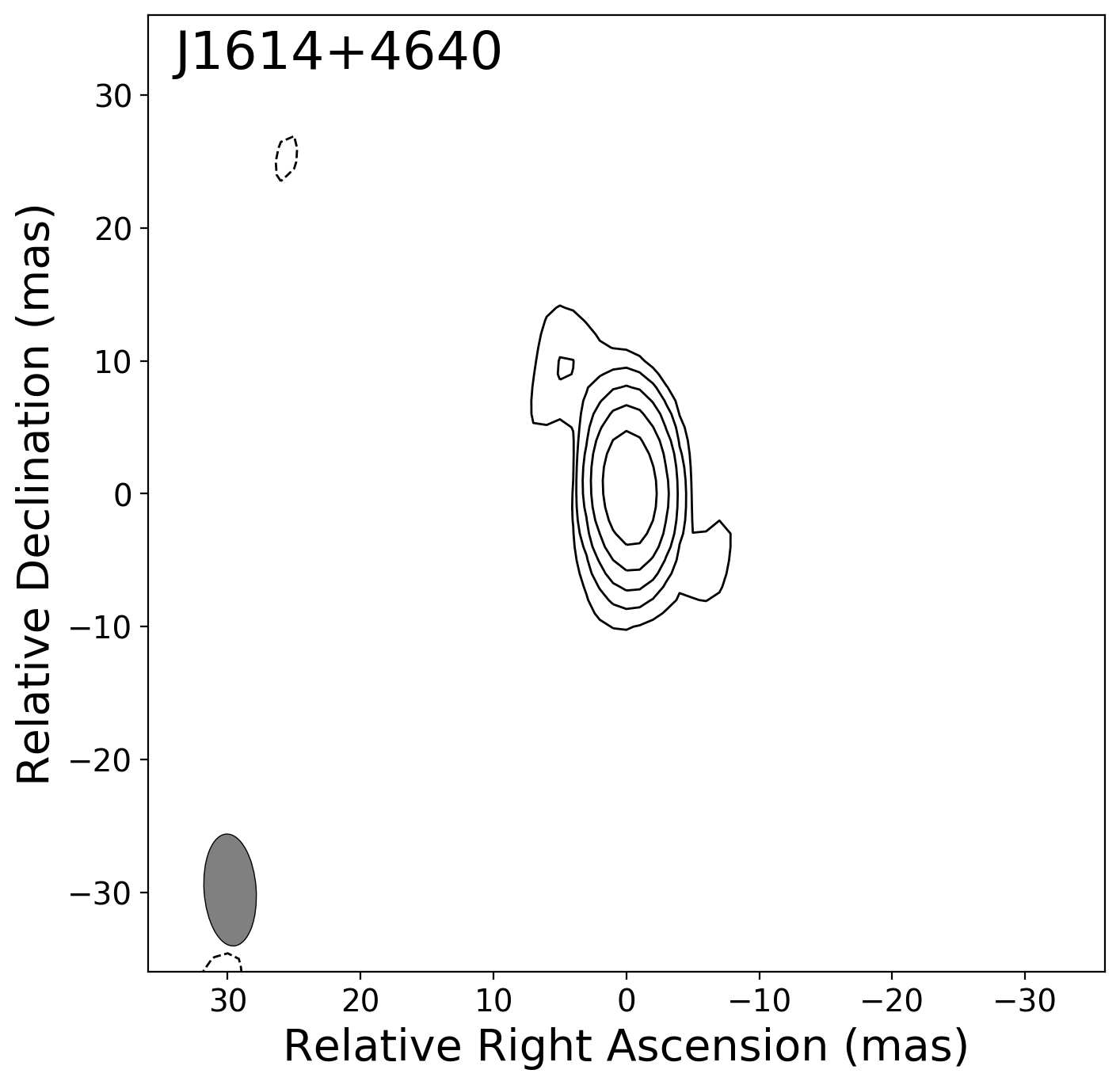}
    \includegraphics[width=0.3\textwidth]{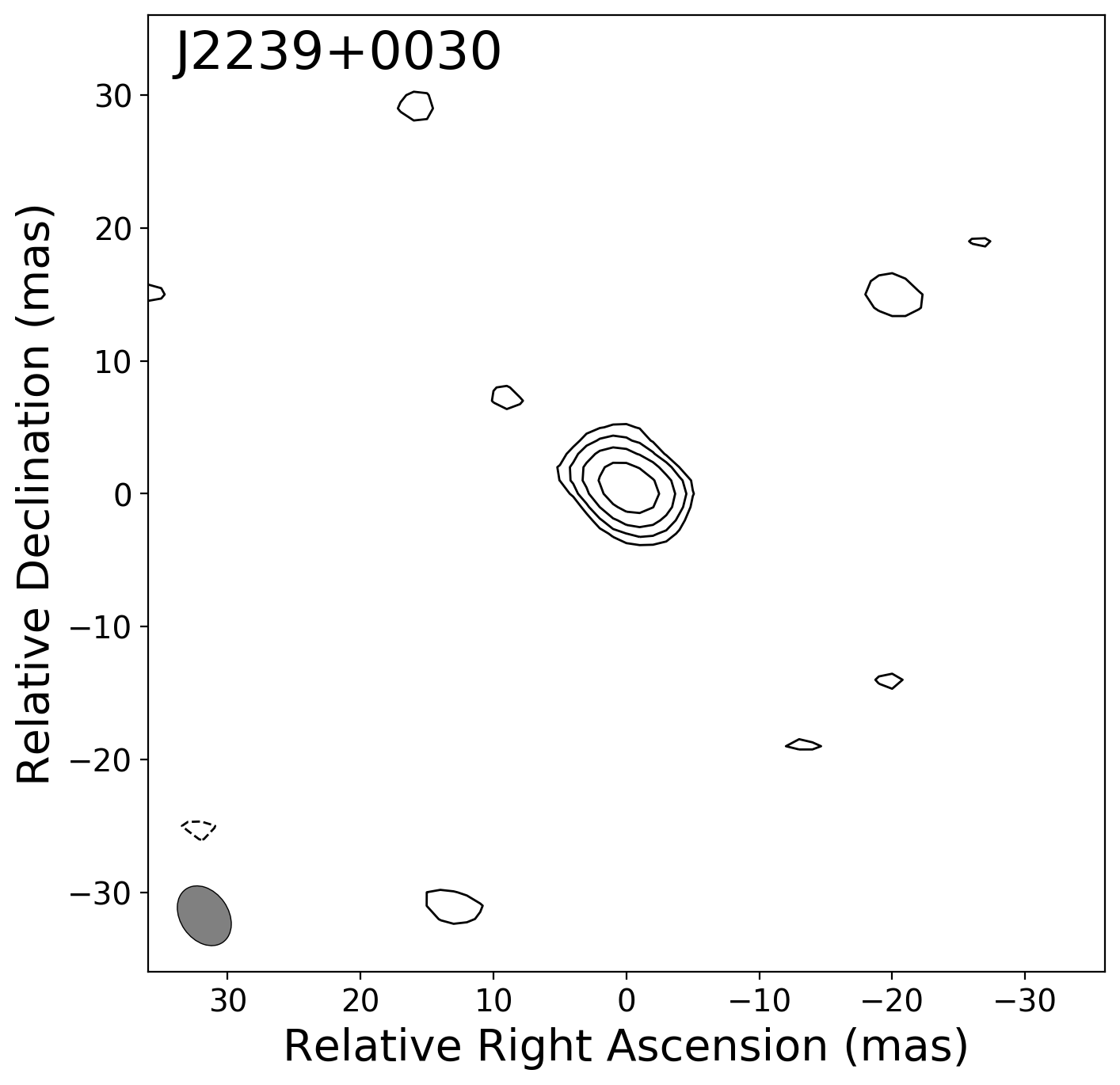}
    \includegraphics[width=0.3\textwidth]{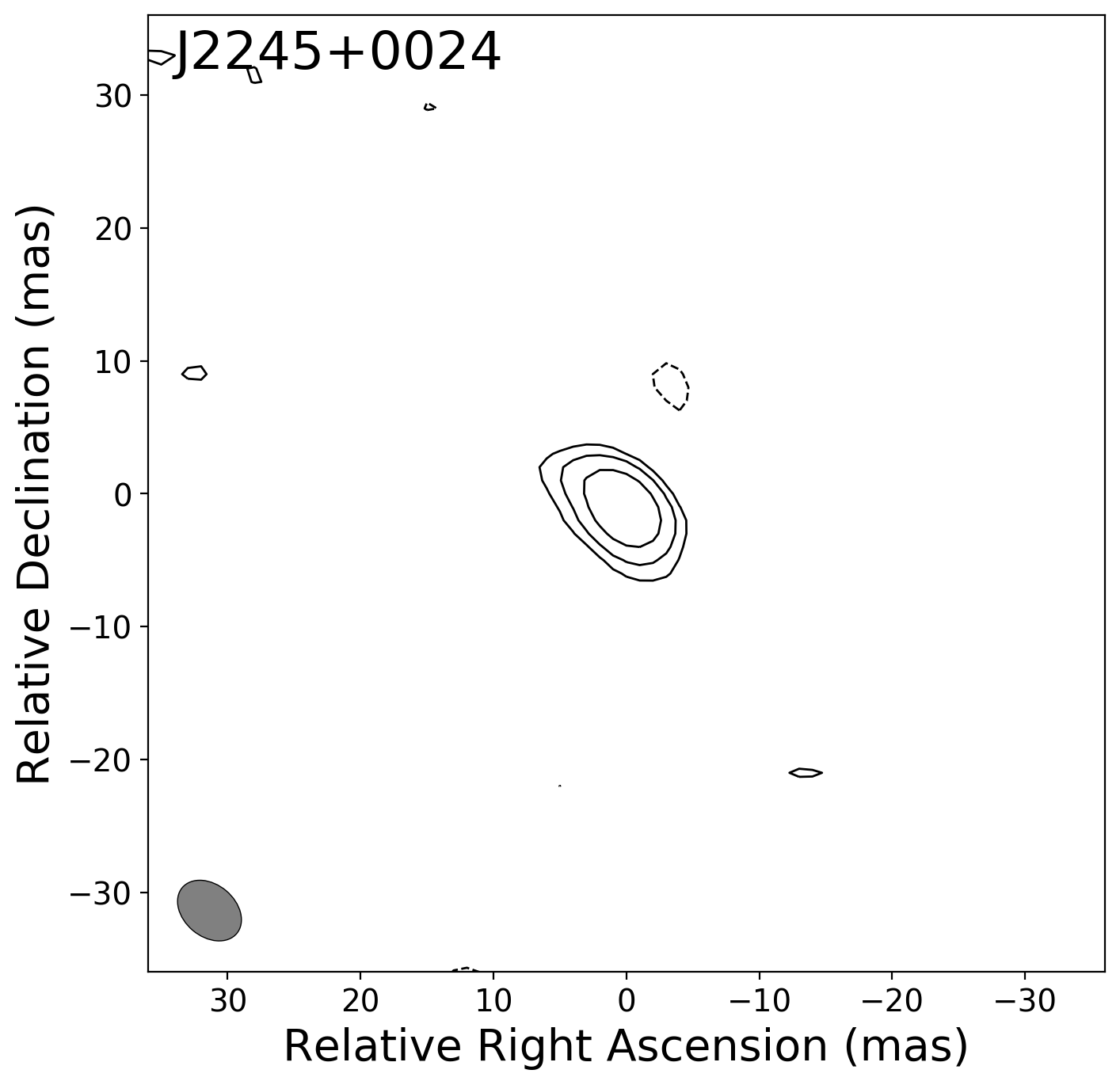}
    \includegraphics[width=0.3\textwidth]{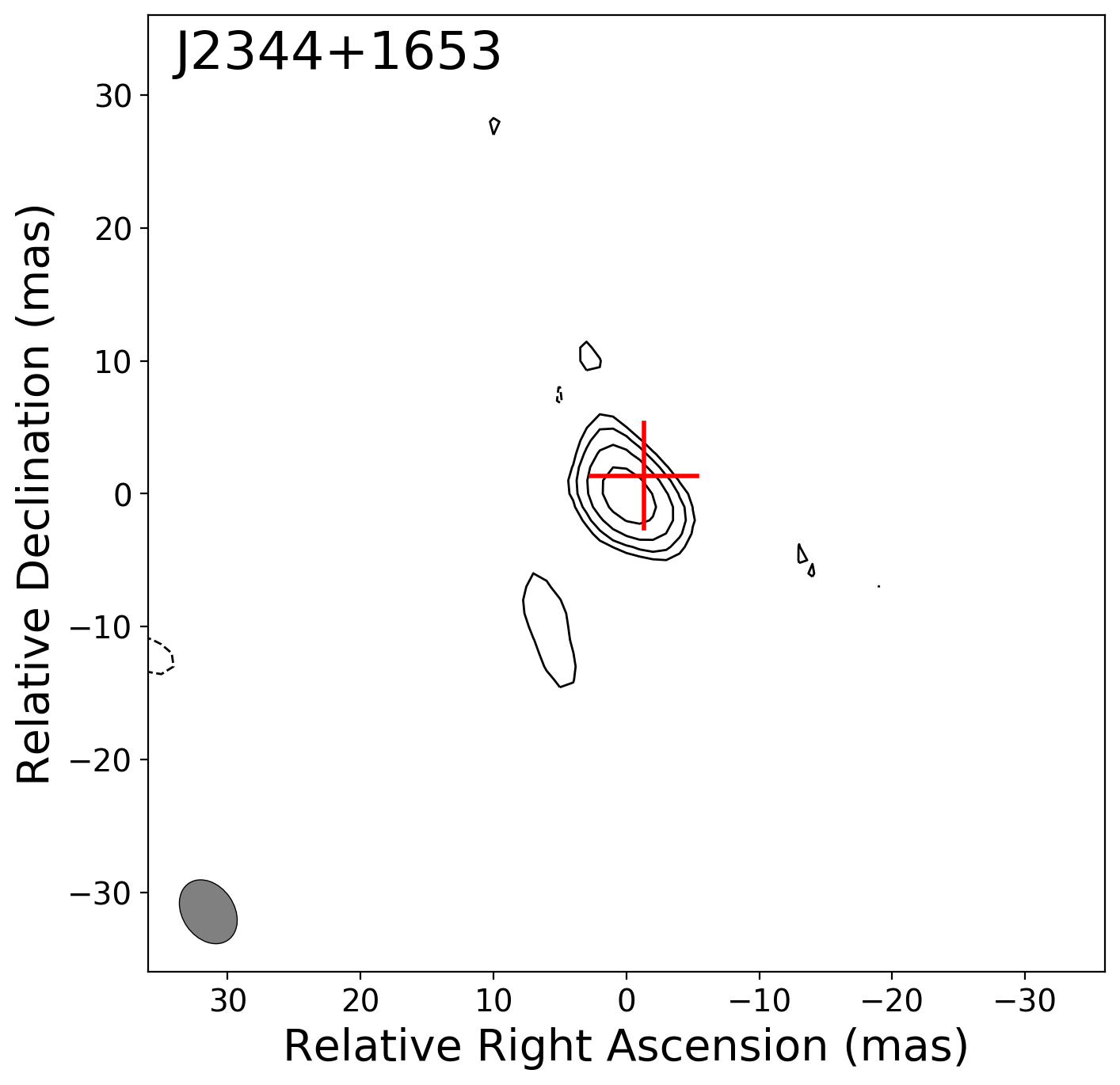}
    \caption{Naturally weighted EVN images at $1.7$~GHz. Red crosses mark the \textit{Gaia} DR3 optical position in the four cases where it is available. The size of the crosses indicates the $3\sigma_{\mathrm{pos}}$ uncertainty (see Sect.~\ref{subsec:gaia}). The lowest contours are drawn at $\pm3$ times the image noise and the positive contours increase by a factor of 2. The restoring beam is shown in the bottom-left corner. Table~\ref{tab:imgparams} contains the image parameters and the coordinates corresponding to the images centres.}
    \label{fig:maps}
\end{figure*}

The naturally weighted 1.7-GHz EVN images are presented in Fig.~\ref{fig:maps}. The image parameters are listed in Table~\ref{tab:imgparams}. Out of the ten target sources, nine were successfully detected as a compact core. The tapering described in Sect.~\ref{reduction} did not reveal extended emission around any of the compact cores in the detected $9$ sources. Only three (namely J1034$+$2033, J1614$+$4640, and J2344$+$1653) of the brighter sources cloud be imaged at higher resolution with applying uniform weighting. J0306$+$1853 was found to be non-detected even after applying different data weighting schemes and tapering. The upper limit to the brightness of this source is $113~\mu$Jy\,beam$^{-1}$ ($5~\sigma$) using the same natural weighting applied for the images in Fig.~\ref{fig:maps} and corrected for $25$\% coherence loss. The parameters derived for the sources are given in Table~\ref{tab:physparams}.

\begin{table*}
    \caption{1.7-GHz VLBI image parameters and source positions.} 
    \label{tab:imgparams}     
    \centering                          
    \begin{tabular}{ccccccc}       
        \hline\hline               
    Source ID  &  RA$_{\mathrm{VLBI}}$ &  Dec$_{\mathrm{VLBI}}$ & Peak intensity & \multicolumn2c{Restoring beam} & $1\sigma$ noise \\  
    &  [$\mathrm{h~min~s}$] & [${^\circ}$~${^\prime}$~$^{\prime\prime}$] & [mJy\,beam$^{-1}$] & [mas $\times$ mas] & [$^\circ$] & [mJy\,beam$^{-1}$] \\
        \hline                      
    J0616$-$1338 & 06 16 24.37784 (0.00016) & $-13$ 38 06.4239 (0.0024) & 1.08 (0.09) & 11.4 $\times$ 3.8 & 71.5 & 0.04   \\ 
    J0741$+$2520 & 07 41 54.71629 (0.00008) & +25 20 29.5594 (0.0012) & 0.81 (0.06) & 4.6 $\times$ 3.6 & 13.4 & 0.03 \\ 
    J0747$+$1153 & 07 47 49.17454 (0.00008) & +11 53 52.4652 (0.0012) & 0.76 (0.06) & 4.6 $\times$ 3.8 & 37.2 & 0.03   \\ 
    J0901$+$1615 & 09 01 32.64577 (0.00008) & +16 15 06.7575 (0.0012) & 1.15 (0.08) & 4.7 $\times$ 3.9 & 34.3 & 0.04  \\ 
    J1034$+$2033 & 10 34 18.64033 (0.00009) & +20 33 00.1247 (0.0013) & 1.68 (0.12) & 5.1 $\times$ 3.7 & 8.8 & 0.05 \\ 
    J1614$+$4640 & 16 14 25.14299 (0.00013) & +46 40 28.9963 (0.0019) & 2.57 (0.10) & 8.4 $\times$ 3.9 & 4.0 & 0.03  \\ 
    J2239$+$0030 & 22 39 07.56146 (0.00008) & +00 30 22.5944 (0.0012) & 0.78 (0.06) & 4.9 $\times$ 3.6 & 34.8 & 0.02  \\ 
    J2245$+$0024 & 22 45 24.27741 (0.00009) & +00 24 14.1389 (0.0013) & 0.45 (0.04) & 5.4 $\times$ 3.9 & 49.8 & 0.02 \\ 
    J2344$+$1653 & 23 44 33.49974 (0.00009) & +16 53 16.5598 (0.0013) & 3.36 (0.23) & 5.2 $\times$ 3.9 & 34.7 & 0.10 \\ 
        \hline                               
    \end{tabular}
\newline
Notes. Col.~1 -- source name; Col.~2 -- right ascension with its uncertainty in seconds (in parentheses); Col.~3 -- declination with its uncertainty in arcseconds (in parentheses); Col.~4 -- image peak intensity; Col.~5 -- restoring beam major and minor axes (FWHM); Col.~6 -- position angle of the restoring beam major axis, measured from north through east; Col.~7 -- $1\sigma$ image noise level.
\end{table*}

The 1.7-GHz redshift-corrected brightness temperatures were calculated following \citep{1982ApJ...252..102C}:
\begin{equation} \label{eq:tb}
    T_{\mathrm{b}} = 1.22 \times 10^{12} \, (1 + z) \frac{S_\nu}{\theta^2 \nu^2} \,\, [\mathrm{K}],
\end{equation}
where $z$ is the redshift, $S_\nu$ the integrated flux density of the core in Jy, $\theta$ the fitted circular Gaussian component diameter (FWHM) in milliarcsecond, and $\nu$ the observing frequency in GHz. To check whether the sources are resolved by the interferometer, we calculated the minimum resolvable size following the formula of \citet{2005AJ....130.2473K}. We found that each of our sources are resolved. It should be noted here that the slightly resolved structure could also be partly an effect of visibility phase decorrelation. 

To investigate the compactness of the sources on the VLBI scale, we adopted the method presented by \citet{2008AJ....136..159L}. They define the compactness as the ratio of the core flux density to the total flux density of the \textsc{clean} components obtained during the imaging, $S_{\mathrm{core}}/S_{\mathrm{clean}}$. Here $S_{\mathrm{core}}$ is equal to the modelled flux density, $S_{\mathrm{1.7}}$, while $S_{\mathrm{clean}}$ is obtained by integrating the flux density contained in the \textsc{clean} components in a 15-mas radius region around the core. The $S_{\mathrm{clean}}$ values are consistent with the modelled flux densities within $10\%$, which indicates that the sources are core-dominated, as seen from their appearance in the VLBI images (Fig.~\ref{fig:maps}).

Another way to characterise the compactness of the sources is to investigate the ratio of their VLBI flux density to the total flux density. However, this indicator gives information about the difference in the radio emission between parsec and kiloparsec scales. We calculated the ratio of the 1.7-GHz VLBI flux density obtained from Gaussian model fits ($S_{\mathrm{1.7}}$) to the total flux density ($S_{\mathrm{VLA,1.7}}$). The latter values are estimated using the derived spectral indices or log-parabolic spectral fits (see more in Section \ref{subsec:spectra}).

Using the formula of \cite{2002astro.ph.10394H}, we calculated the core and total 1.7-GHz redshift-corrected monochromatic radio powers, $P_{\mathrm{1.7, core}}$ and $P_{\mathrm{1.7, total}}$, respectively:
\begin{equation} \label{eq:power}
    P_{\nu} = 1.20 \times 10^{20} \, D_{\mathrm{L}}^2 S_{\nu} (1+z)^{-1-\alpha} \,\, [\mathrm{W}\,\mathrm{Hz}^{-1}]. 
\end{equation}
In this equation, $D_\mathrm{L}$ refers to the source luminosity distance in Megaparsec unit, $\alpha$ is the radio spectral index (see Table~\ref{tab:sedparams}), and $S_{\nu} = S_{\mathrm{1.7}}$ for $P_{\mathrm{1.7, core}}$ (assuming the same spectral index for the VLBI-scale emission as derived from the total flux densities), while $S_{\nu} = S_{\mathrm{VLA,1.7}}$ for $P_{\mathrm{1.7,  total}}$. We applied $\alpha=0$ to sources with peaked spectrum because the spectral shape in the frequency range where the K-correction is applied is close to flat (see Fig.~\ref{fig:sed}).

\begin{table*}
    \caption{Derived physical parameters of the sources.} 
    \label{tab:physparams}     
    \centering                          
    \begin{tabular}{ccccccccc}       
        \hline\hline               
    Source ID & $S_\mathrm{1.7}$ & $\theta_\mathrm{1.7}$ & $T_\mathrm{b,1.7}$  & $S_{\mathrm{clean}}$ & $S_{\mathrm{VLA,1.7}}$ & $S_{\mathrm{1.7}}/S_{\mathrm{VLA,1.7}}$ & $P_{\mathrm{1.7, core}}$ & $P_{\mathrm{1.7, total}}$\\
     & [mJy] & [mas] & [$10^8$ K] & [mJy] &  [mJy] & & [10$^{26}$ W~Hz$^{-1}$] & [10$^{26}$ W~Hz$^{-1}$]\\
        \hline 
    J0306$+$1853 & $<0.11^{*}$ & ... & ... &  ... & 0.23 (0.01) & ... & ... & 0.57 (0.08) \\
    J0616$-$1338 & 1.86 (0.27) & 2.3 (0.1) & 10.6 (2.0) & 1.86 &  ... & ... & 3.86 (0.73) & ... \\ 
    J0741$+$2520 & 2.69 (0.36) & 8.5 (0.5) & 1.0 (0.2) & 2.54 & 3.33 (0.41) & 0.81 (0.21) & 1.24 (0.22) & 1.54 (0.26)  \\ 
    J0747$+$1153 & 1.79 (0.24) & 4.0 (0.2) & 3.1 (0.6) & 1.81 & ... & ...  & 1.66 (0.29) & ...\\ 
    J0901$+$1615 & 3.06 (0.39) & 8.4 (0.5) & 1.3 (0.2) & 3.19 & 3.68 (0.17) & 0.82 (0.15) & 3.96 (0.70) & 4.75 (0.62) \\ 
    J1034$+$2033 & 3.46 (0.46) & 3.2 (0.2) & 9.2 (1.6) & 3.56  & 4.18 (0.14) & 0.83 (0.14) & 1.51 (0.27) & 1.83 (0.22) \\ 
    J1614$+$4640 & 3.45 (0.35) & 1.5 (0.1) & 41.7 (4.9) & 3.96 & 2.95 (0.41) & 1.23 (0.29) & 1.73 (0.27) & 1.41 (0.29) \\ 
    J2239$+$0030 & 1.36 (0.18) & 2.2 (0.1) & 7.7 (1.4) & 1.56 & 1.06 (0.21) & 1.28 (0.42) & 1.06 (0.19) & 0.83 (0.19) \\ 
    J2245$+$0024 & 0.76 (0.12) & 4.0 (0.3) & 1.3 (0.3) & 0.76 & 1.31 (0.14) & 0.58 (0.16) & 0.35 (0.07) & 0.60 (0.09) \\ 
    J2344$+$1653 & 7.03 (0.89) & 1.9 (0.1) & 53.0 (8.9) & 6.67 & 14.13 (0.47) & 0.50 (0.08) & 9.31 (1.61) & 18.71 (2.27) \\ 
        \hline                               
    \end{tabular}
\newline
Notes. $^{*}$ non-detection, an upper limit to the peak brightness is given in mJy\, beam$^{-1}$; the errors are shown in parentheses; Col.~1 -- source name; Col.~2 -- 1.7-GHz fitted core flux density, with errors shown in parentheses; Col.~3 -- fitted FWHM of the core component; Col.~4 -- redshift-corrected 1.7-GHz brightness temperature; Col.~5 -- \textsc{clean}ed flux density integrated within the central 15-mas radius area; Col.~6 -- 1.7-GHz total flux density estimated using the derived spectral indices and log-parabolic coefficients (see Table~\ref{tab:sedparams}); Col.~7 -- ratio of the measured VLBI and the calculated total 1.7-GHz flux densities; Col.~8 -- 1.7-GHz monochromatic radio power for the core and its uncertainty; Col.~9 -- 1.7-GHz total monochromatic radio power and its uncertainty.
\end{table*}

\section{Discussion} \label{discussion}

\subsection{The origin and the nature of the radio emission}

The high detection rate of compact parsec-scale jets in RQQs in this study (three sources out of four were detected) contrasts with the low detection rates seen in low-redshift RQQ samples \citep{2018MNRAS.476.3478B,2021MNRAS.500.4749B,2023MNRAS.518...39W,2023MNRAS.525.6064W}. This potentially suggests an interesting evolution in the jet properties and/or launch mechanisms of RQQs with cosmic time. Selection effects still could play a role in this contradiction. It should be noted when the results are compared with the low-redhsift samples that the rest-frame frequencies for our 1.7-GHz measurements at $5 < z < 6$ are equal to $\sim 10-12$~GHz. The difference between the rest-frame frequencies could cause bias in the comparison. Nevertheless, the detected sources may represent the tip of the iceberg, the brightest, most luminous of the high-redshift RQQ population. Further deeper VLBI observations probing fainter sources would test whether the high detection rate holds. It is possible that the RLQ/RQQ dichotomy itself becomes increasingly blurred at high redshifts, if a larger fraction of AGN have relativistic jets due to different accretion physics or environmental conditions in the early Universe.

The sources detected in our sample appear to be highly core-dominated. They seem to lack extended emission due to the losses coming from scattering with cosmic microwave background (CMB) photons \citep{2015MNRAS.452.3457G}, or perhaps the parsec-scale jets are beamed towards the observer. It is also possible that these radio AGN are very young, powerful, and intrinsically small sources \citep{2021A&ARv..29....3O}.
The derived brightness temperatures (see Table~\ref{tab:physparams}) all exceed $T_{\mathrm{b}} = 10^6$~K by at least two orders of magnitude, clearly indicating non-thermal radio emission related to AGN activity and ruling out star formation in the host galaxy as the primary source of radio emission \citep{1992ARA&A..30..575C,2000ApJ...530..704K,2011A&A...526A..74M}. The AGN-associated activity may include jets, outflows and coronal emission \citep{2019NatAs...3..387P}. The compactness and the derived $T_{\mathrm{b}} \sim 10^{8-9}$~K values strengthen the case for the presence of plasma jets in our sources. Jetted radio sources often show relativistically enhanced radio emission, depending on the jet inclination to the line of sight and the bulk speed of the plasma. The jet emission is considered Doppler-boosted when $T_{\mathrm{b}}$ exceeds the equipartition value of $T_{\mathrm{b,eq}} \approx 5 \times 10^{10}$~K \citep{1994ApJ...426...51R}. Having the highest measured $T_{\mathrm{b}} = (5.3 \pm 0.9) \times 10^9$~K in our sample, the radio emission in none of these sources appears Doppler-boosted, suggesting their jets are not aligned close to the line of sight. 
However, VLBI imaging at higher observing frequency and thus with better angular resolution might be able to reveal higher brightness temperatures in these sources, depending on the shape of the radio spectrum.

The parsec-scale morphology of the radio structures in this sample of nine VLBI-detected AGN is generally similar to what is seen in other high-redshift VLBI samples \citep[e.g.][]{2016MNRAS.463.3260C,2022ApJS..260...49K}. There is no clear sign of extended emission beyond a few mas radius or other jet component in addition to the compact core. The only exception is J2344$+$1653, where the $S_{\mathrm{1.7}}/S_{\mathrm{VLA,1.7}}$ ratio is $0.50 \pm 0.08$. It suggests the presence of a diffuse ($\sim 0.1-1\arcsec$) emission around the compact core which is resolved out by the EVN. This could also apply for J2245$+$0024, but the uncertainties are larger for that source (Table~\ref{tab:physparams}).

\subsection{Gaia optical astrometric positions} \label{subsec:gaia}

The \textit{Gaia} astrometric space mission provides the most accurate optical positions available nowadays for AGN \citep{2016A&A...595A...1G, 2023A&A...674A...1G}. Thanks to the sensitivity of \textit{Gaia}, its catalogue contains even some faint distant AGN. Comparing their coordinates with the radio positions derived from VLBI observations which have similar (milliarcsecond or sub-milliarcsecond) accuracy, additional information on the nature of the investigated sources could be revealed. The optical position of a quasar is generally linked to the thermal emission originating from the accretion disk \citep{2017A&A...598L...1K,2019ApJ...871..143P}, or sometimes to synchrotron emission of bright optical jets \citep[e.g.][]{2024A&A...684A.202L}.
For high-redshift quasars like our target sources, the redshift is determined using broad Ly${\alpha}$ lines formed in the broad-line region (BLR) in the vicinity of the central black hole \citep{2006LNP...693...77P}. In case of a dominant non-thermal optical emission from the relativistic jet, the thermal component from the BLR would be suppressed. Therefore, the detection of the broad emission lines in the optical spectra of high-redshift quasars favours the thermal origin of their optical emission.
On the other hand, the radio emission corresponds to the synchrotron self-absorbed base of the jet (i.e. the radio core) with $\tau_{\nu}$ = 1 optical depth at the given frequency $\nu$, or possibly a compact hotspot associated with a shock front between the jet and the surrounding medium. The radio core is usually the brightest and most compact region of the jet and can be pinpointed with VLBI observations \citep[e.g.][]{1997AJ....114.2284F}. Up to a few milliarcsecond offset between radio and optical AGN positions can occur and its position angle is found to statistically coincide with the VLBI jet direction \citep{2019MNRAS.482.3023P}. For faint radio sources like the typical high-redshift quasars, revealing a significant \textit{Gaia}--VLBI positional mismatch exceeding the typical values could help constraining the source classification and the nature of compact radio emission, as it would suggest an extended structure with a quasar jet misaligned with respect to the line of sight.

Out of the ten targets in our sample, five (namely J0306$+$1853, J0741$+$2520, J0747$+$1153, J1034$+$2033, and J2344$+$1653) have been detected by \textit{Gaia}. We looked for counterparts for the remaining five sources in the \textit{Gaia} source catalogue, but there were no matches within $\sim 1\arcsec$ angular distance. The \textit{Gaia} positions are marked with red crosses in Fig.~\ref{fig:maps}. The uncertainties of the optical positions from the \textit{Gaia} DR3 \citep{2023A&A...674A...1G} are within 3~mas. The astrometric excess noise is only significant in the case of J0747$+$115, where we added it to the formal position error in quadrature. Since the quasar J0306$+$1853 remained undetected with the EVN, we could not derive its accurate radio position to be compared with the \textit{Gaia} position. 

The VLBI astrometric positions determined for our targets by phase-referencing to nearby calibrator sources can be considered as linked to the ICRF \citep{2020A&A...644A.159C}. Our measured \textit{Gaia}--VLBI positional offsets are ranging between $1.3-1.8$~mas. However, ICRF positions refer to X band ($8.4$~GHz), while our phase-referenced observations were carried out at L band ($1.7$~GHz). The frequency-dependent core shift in a synchrotron self-absorbed jet can amount $\sim 1$~mas between these frequencies \citep[e.g.][]{2011A&A...532A..38S}. Moreover, according to \citet{2009A&A...505L...1P}, there could be up to $\sim 0.2$-mas offset between the phase-referenced and group-delay positions along the jet direction, caused by opacity effects on the emission at the jet base.
An offset can be considered significant if the optical position differs by more than $3\sigma_{\mathrm{pos}}$, where $\sigma_{\mathrm{pos}} = \sqrt{\sigma_\mathrm{VLBI}^2+\sigma_\mathrm{Gaia}^2}$, the square root of the sum of squares of the VLBI and the \textit{Gaia} positional uncertainties. Based on this criterion, we find the optical--radio offsets insignificant. This in fact indicates
that the detected radio features could be related to the inner jet, located close to the central engine.

\subsection{Continuum radio spectra}\label{subsec:spectra}

\begin{figure*}[!ht]
    \centering
    \includegraphics[width=0.3\textwidth]{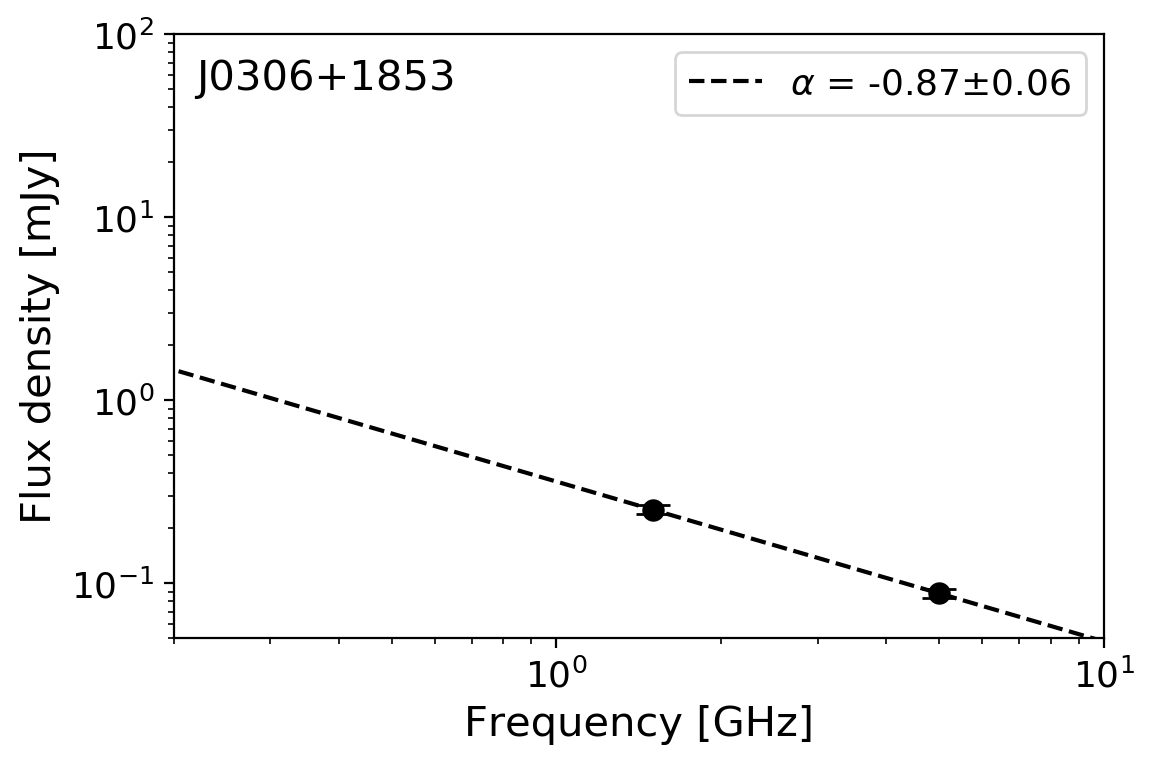}
    \includegraphics[width=0.3\textwidth]{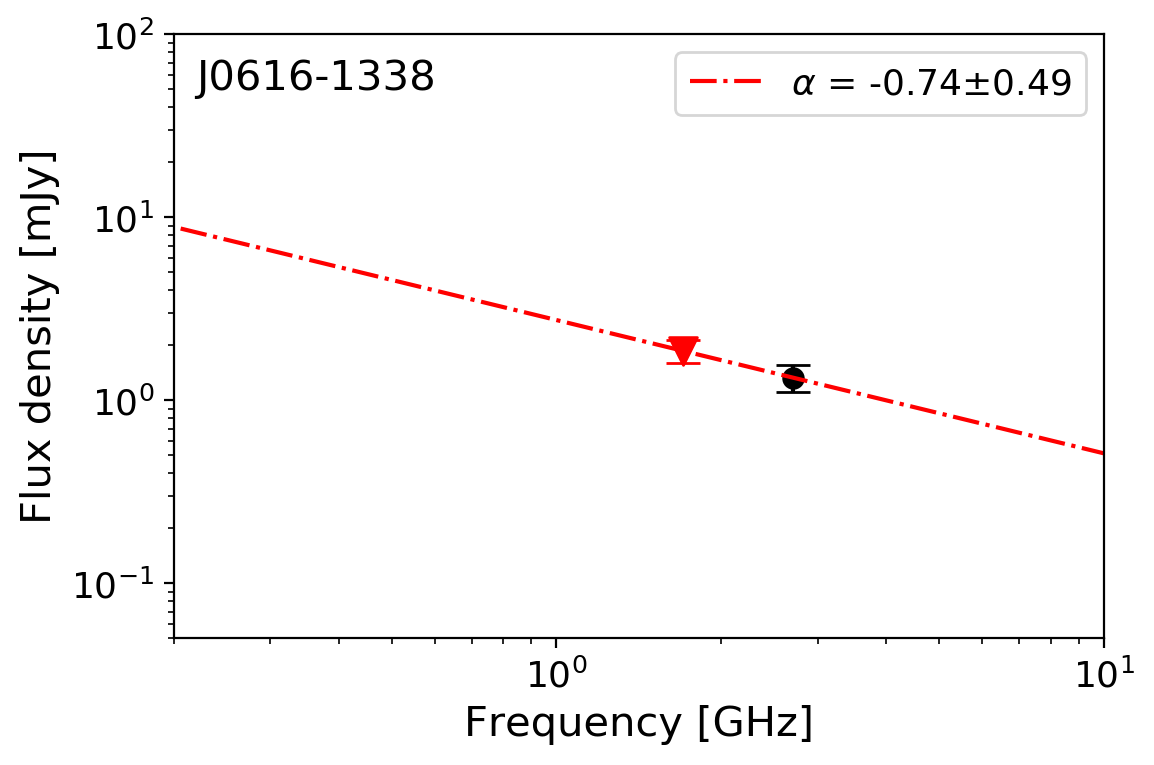}
    \includegraphics[width=0.3\textwidth]{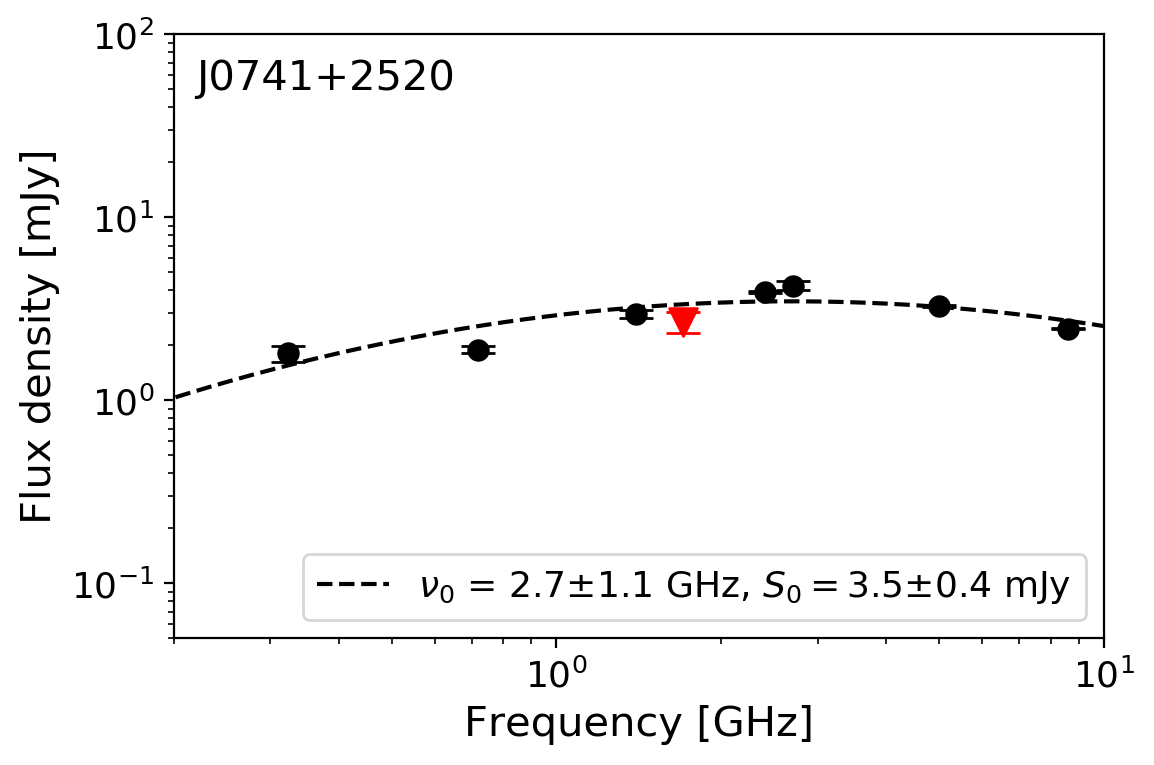}
    \includegraphics[width=0.3\textwidth]{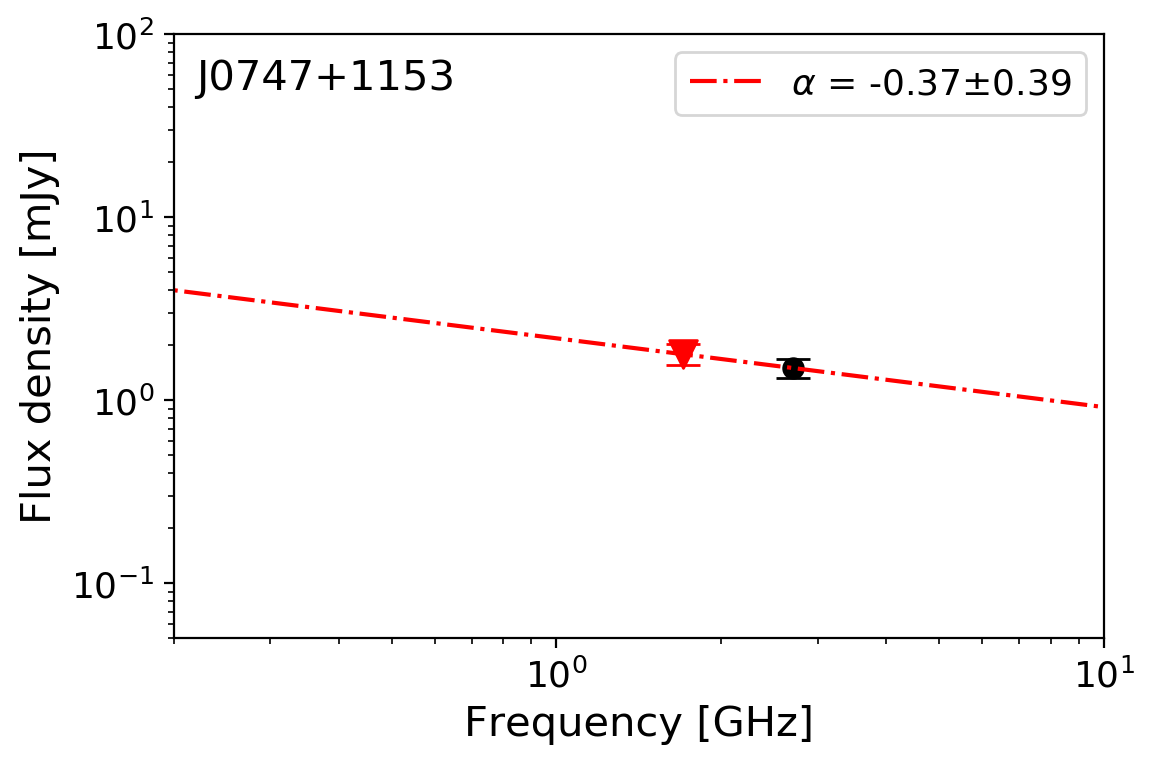}
    \includegraphics[width=0.3\textwidth]{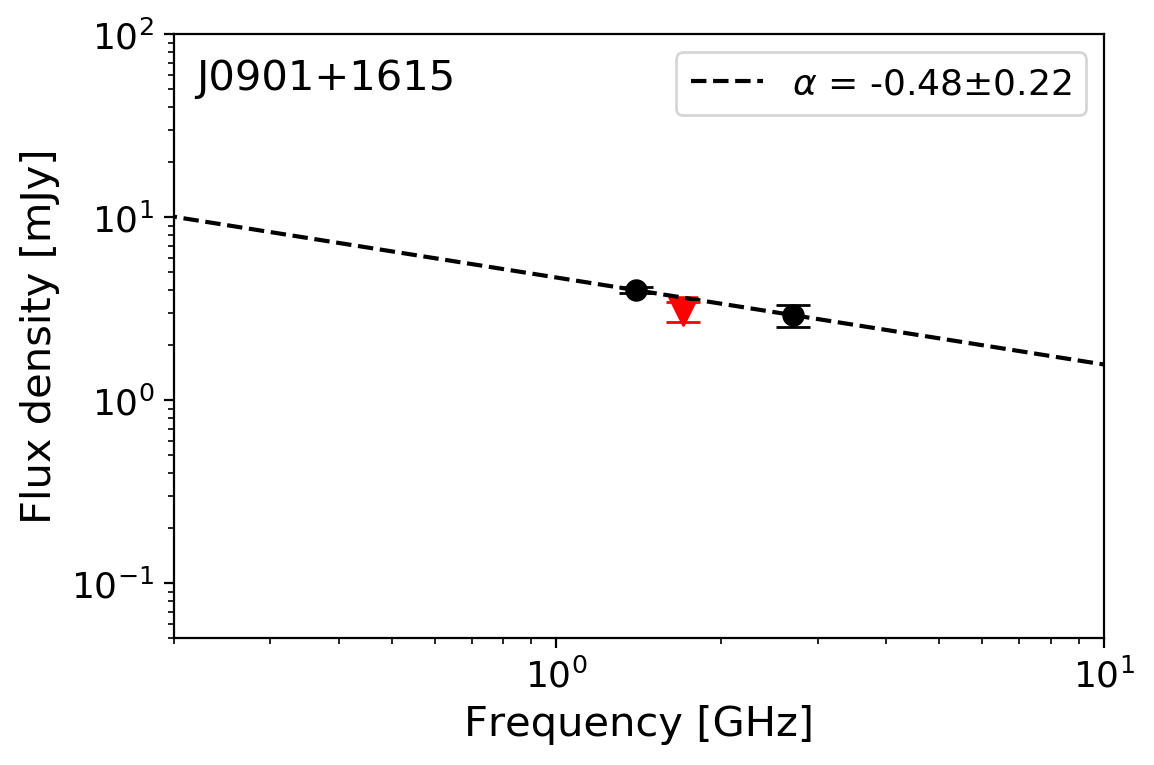}
    \includegraphics[width=0.3\textwidth]{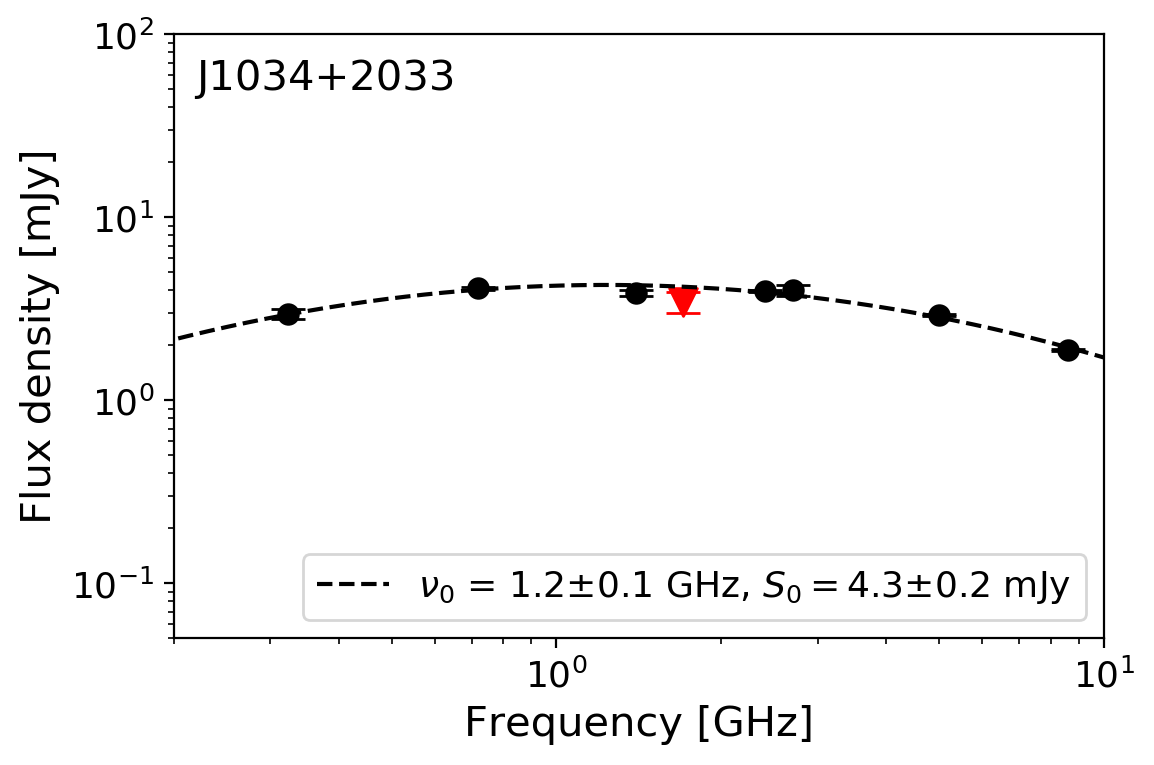}
    \includegraphics[width=0.3\textwidth]{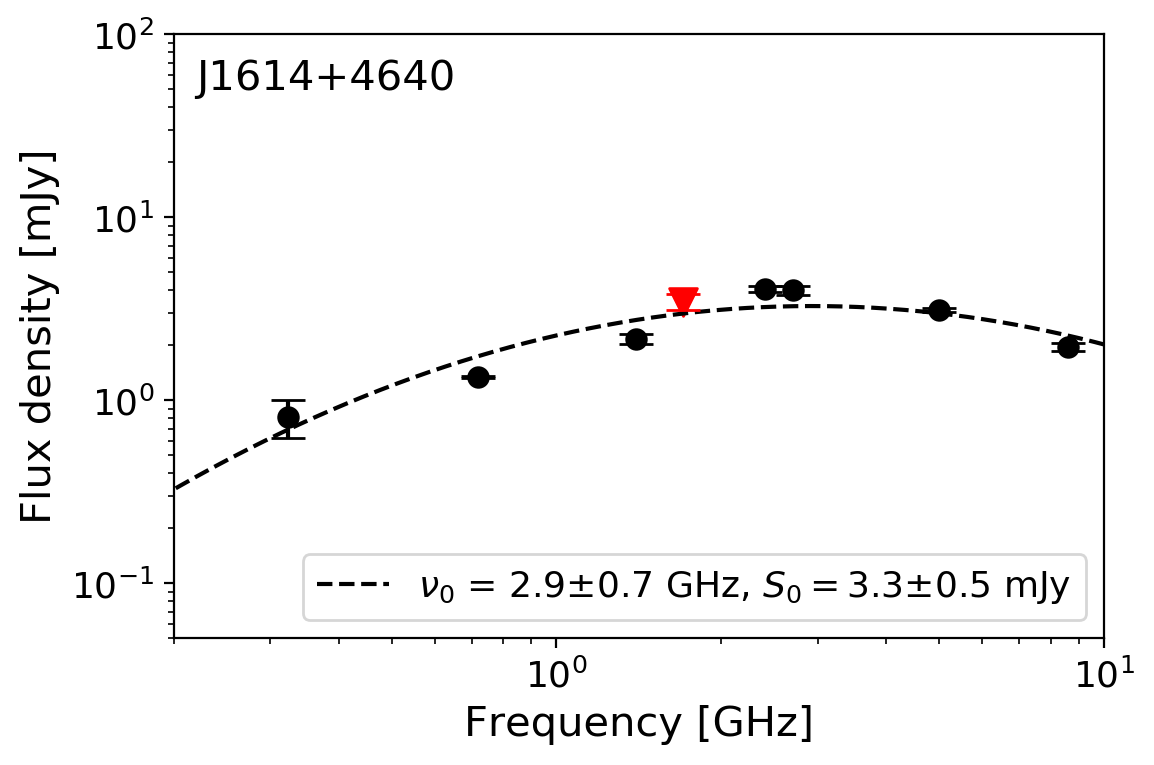}
    \includegraphics[width=0.3\textwidth]{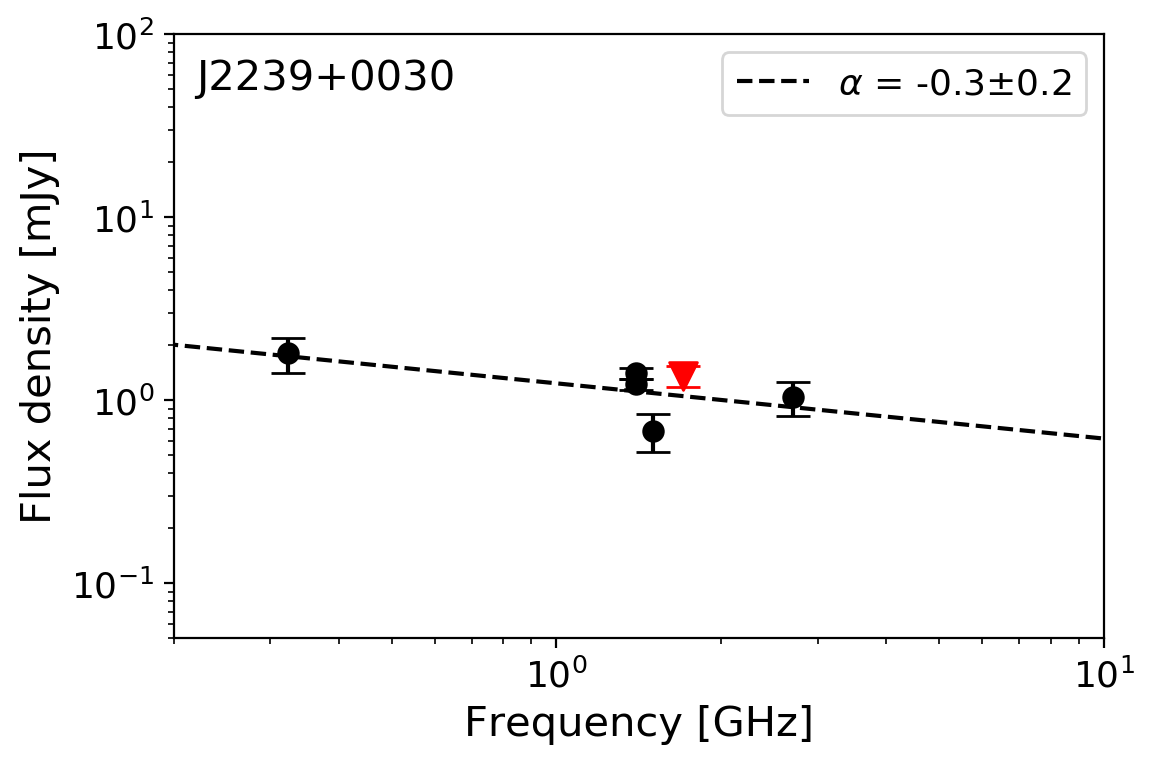}
    \includegraphics[width=0.3\textwidth]{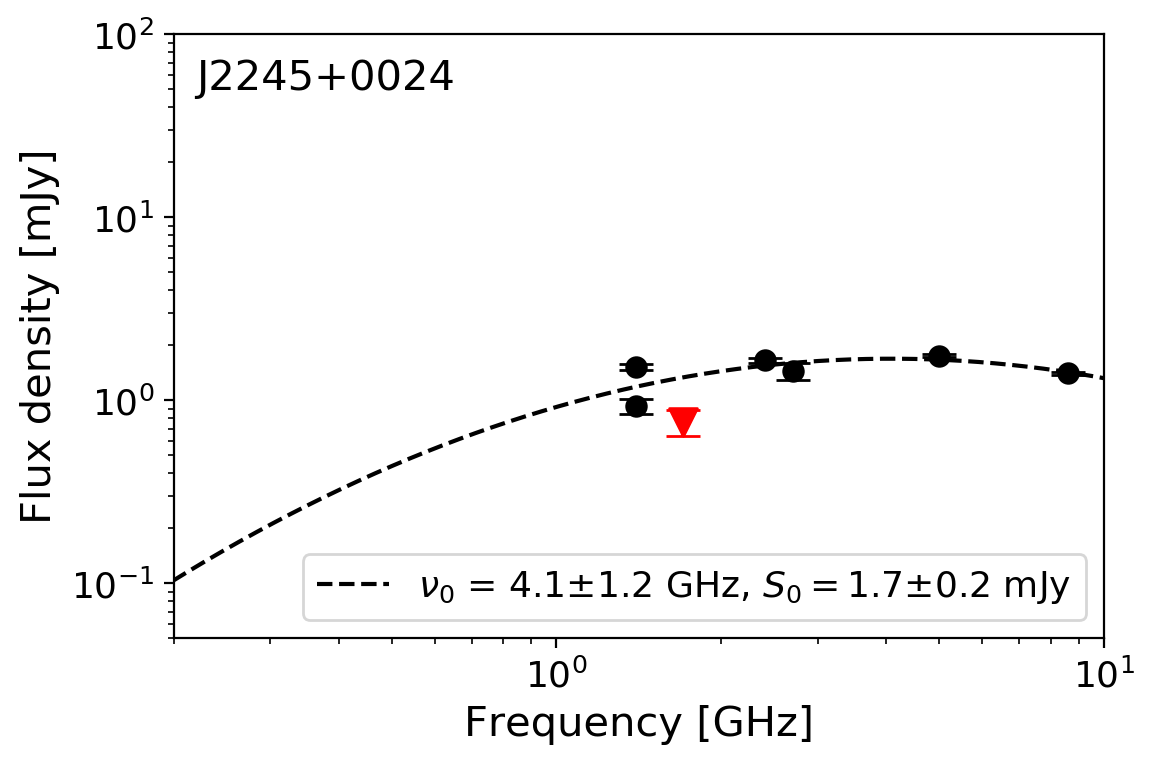}
    \includegraphics[width=0.3\textwidth]{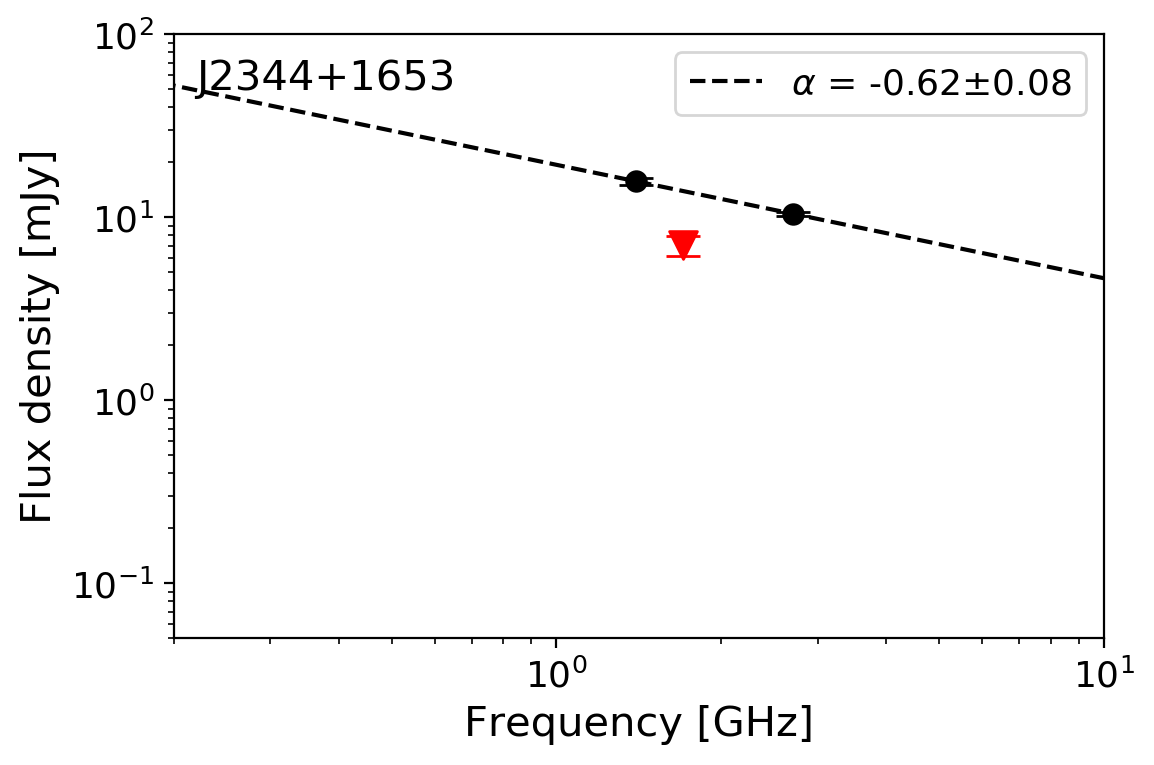}
    \caption{Radio continuum spectra of the ten sources in our sample. Black points indicate total flux density data from single-dish and low-resolution interferometric  measurements. The black dashed lines show the best-fit spectrum (power-law or log-parabolic). The parameters of the spectral fits are given in the insets and listed in Table~\ref{tab:sedparams}. 
    The red data points are the 1.7-GHz EVN flux density measurements presented in this paper. The red dashed-dotted lines for J0616$-$1338 and J0747$+$1153 show the power-law function fitted to the only available two flux density data points (VLASS and EVN), with the assumption that all the radio emission is confined to a small spatial scale corresponding to the compact VLBI component.}
    \label{fig:sed}
\end{figure*}

We collected the available single-dish and low-resolution radio interferometric observations from the literature for the targets, in order to build their total flux density spectra. The investigated quasars are faint ($S_{\mathrm{2.7}} <  11$~mJy, Table~\ref{tab:sample}), thus they are often below the detection thresholds of shallow surveys. The main sources of the flux density measurements are the GMRT (Giant Metrewave Radio Telescope) and uGMRT (upgraded GMRT) observations of radio-loud quasars at $z > 5$ \citep{2020AA...641A..85S,2022AA...659A.159S} and various surveys carried out with the VLA (Table~\ref{tab:sedparams}). The continuum spectra for the ten sources in our sample are shown in Fig.~\ref{fig:sed}. There are two objects, J0616$-$1338 and JJ0747$+$1153, where only the 2.7-GHz VLASS measurements are available. For the rest of the sample, we fitted the spectral data points to characterise their continuum spectrum (black-coloured fits in Fig.~\ref{fig:sed}). In four cases, we assumed a power-law radio spectrum $S \propto \nu^{\alpha}$, where $S$ is the flux density, $\nu$ the frequency, and $\alpha$ the spectral index. Power-law spectra are called steep if $\alpha < -0.5$, flat if $-0.5 \leq \alpha \leq 0$, and inverted if $\alpha > 0$. In the case of four sources (J0741$+$2520, J1034$+$2033, J1614$+$4640, and J2245$+$0024) with an apparently curved spectral shape, we found that the choice of a log-parabolic function in the form of $\log S = a (\log \nu - \log~\nu_0)^2 + b$ resulted in lower $\chi^2$ values than using the power-law function. In the formula above, $\nu_0$ is the turnover frequency corresponding to the peak flux density $S_0$, while $a$ and $b$ are numerical constants \citep[e.g.][]{2017MNRAS.467.2039C}. The log-parabolic fit can be characterised by the $\nu_0$ and $S_0$ values. Given that some sources in our sample lack low-frequency data points, it is possible that the their spectra are in fact also peaked. Peaked continuum radio spectra seem to be common for bright $z>3$ radio quasars \citep{2021MNRAS.508.2798S}. Also, \citet{2022AA...659A.159S} found that $z > 5$ radio-loud quasars tend to show evidence for spectral turnover at rest-frame frequencies of $\sim 1-50$\,GHz. 

In Table~\ref{tab:sedparams}, we give the flux densities and the derived spectral parameters, i.e. the spectral indices in the case of power-law fits, and the peak frequency ($\nu_0$) and flux density ($S_0$) values for log-parabolic fits. Out of the four sources whose flux densities versus frequency are best described with a power-law function, two show flat and two steep spectrum. The four peaked spectra have their maxima between $1-5$~GHz which corresponds to $\sim 6-30$~GHz in the rest frame of the sources. The fitted $\nu_0$ and $S_0$ values are consistent within the uncertainties with the results of \citet{2022AA...659A.159S} who also found these sources having peaked spectra. The difference between our and their analyses is that we also included the 1.4-GHz FIRST and 2.7-GHz VLASS data points.  

Deriving spectral indices of the compact VLBI components would require VLBI measurements in at least two different frequency bands taken preferably at around the same time, to avoid possible effects of variability. If the source is compact enough that the VLBI flux density is comparable to the low-resolution measurements, the VLBI flux density can be included in the total flux density spectral analyses to help constrain the fit. We found the compactness of J0616$-$1338 and J0747$+$1153 to be close to unity (Table~\ref{tab:physparams}), therefore we used our 1.7-GHz EVN measurements together with the 2.7-GHz VLASS data to derive their spectral index (red-coloured fits in Fig.~\ref{fig:sed}). J0616$-$1338 has steep spectrum with $\alpha^{\mathrm{2.7~GHz}}_{\mathrm{1.7~GHz}} = -0.72\pm0.49$, although the large uncertainty allows for a flat spectrum as well. J0747$+$1153 has formally a flat spectrum, but the uncertainty is also large because there are only two spectral points available, at relatively close frequencies.

\begin{table*}[!ht]
    \footnotesize
    \caption{Total flux density data used collected from the literature.} 
    \label{tab:sedparams}     
    \centering                          
    \begin{tabular}{cccccccccc}       
        \hline\hline               
    Source ID & $\alpha$ & $S_\mathrm{323~MHz}$ & $S_\mathrm{720~MHz}$  & $S_\mathrm{1.4~GHz}$ & $S_\mathrm{2.4~GHz}$ & $S_\mathrm{2.7~GHz}$ & $S_\mathrm{5~GHz}$ & $S_\mathrm{8.6~GHz}$  \\
     & & [mJy]  & [mJy] & [mJy] & [mJy] & [mJy] & [mJy] & [mJy] \\
        \hline 
    J0306$+$1853 & $-0.87$ (0.08) & & & 0.25 (0.01)$^1$ & & & 0.09 (0.01)$^1$ & \\
    J0616$-$1338 & $-0.74$ (0.49)$^*$ & & & & & 1.33 (0.22) & & \\
    J0741$+$2520 & $S_0 =$ 3.5 (0.4) & 1.80 (0.19)$^4$ & 1.89 (0.08)$^5$ & 2.97 (0.14)$^3$ & 3.27 (0.03)$^5$ & 4.23 (0.24)$^2$  & 3.89 (0.05)$^5$ & 2.46 (0.02)$^5$ \\
                 & $\nu_0 =$ 2.7 (1.1) & & & & & & & \\
    J0747$+$1153 & $-0.37$ (0.39)$^*$ & & & & & 1.50 (0.17)$^2$ & & \\
    J0901$+$1615 & $-0.48$ (0.22) & & & 3.99 (0.15)$^3$ & & 2.92 (0.41)$^2$ & & \\
                 & & &              & & & & & \\
    J1034$+$2033 & $S_0 =$ 4.3 (0.2) & 2.97 (0.19)$^4$ & 4.08 (0.07)$^5$ & 3.85 (0.14)$^3$ & 3.94 (0.06)$^5$ & 3.99 (0.29)$^2$ & 2.92 (0.04)$^5$ & 1.88 (0.03)$^5$ \\
                 & $\nu_0 =$ 1.2 (0.1) & & & & & & & \\
    J1614$+$4640 & $S_0 =$ 3.3 (0.5) & 0.81 (0.19)$^4$ & 1.34 (0.01)$^5$ & 2.16 (0.14)$^3$ & 4.06 (0.16)$^5$ & 4.00 (0.23)$^2$  & 3.13 (0.08)$^5$ & 1.96 (0.10)$^5$ \\
                 & $\nu_0 =$ 2.9 (0.7) & & & & & & & \\
    J2239$+$0030 & $-0.31$ (0.19) & 1.80 (0.39)$^4$ & & 1.40 (0.10)$^3$ & & 1.04 (0.22)$^2$ & \\
                 & & & & 1.22 (0.08)$^6$ & & & & \\
                 & & & & 0.68 (0.16)$^7$ & & & & \\
    J2245$+$0024 & $S_0 =$ 1.7 (0.2) & & & 0.93 (0.09)$^3$ & 1.65 (0.06)$^5$ & 1.44 (0.15)$^2$  & 1.75 (0.03)$^5$ & 1.40 (0.03)$^5$ \\
                 & $\nu_0 =$ 4.1 (1.2) & & & 1.52 (0.06)$^6$ & & & & \\
    J2344$+$1653 & $-0.62$ (0.08) & & & 15.7 (0.7)$^8$ & & 10.12 (0.24)$^2$ & & \\ 
        \hline                               
    \end{tabular}
\newline
Notes. $^*$ two-point spectral index derived from 1.7-GHz VLBI and 2.7-GHz VLASS flux density measurements; Col.~1 -- source designation; Col.~2 -- fitted power-law spectral index; where log-parabolic function was fitted, we give values $S_0$ in mJy and $\nu_0$ in GHz; Col.~3--9 -- flux densities at the various frequencies and their uncertainties collected from the literature; numbers in the upper indices are references to the measurements: 1: \citet{2021AA...655A..95S}, 2: VLASS, 3: FIRST, 4: \citet{2020AA...641A..85S}, 5: \citet{2022AA...659A.159S}, 6: \citet{2011AJ....142....3H}, 7: \citet{2016MNRAS.460.4433H}, and 8: \citet{1998AJ....115.1693C}.
\end{table*}

\subsection{Variability}

Variability is an indicator of relativistic jet or coronal emission for radio-quiet AGN \citep[e.g.][]{2023MNRAS.525.6064W} and of powerful jets in high-redshift radio-loud AGN \citep[e.g.][]{2024Galax..12...25S}. Due to the lack of measurements performed at the same frequencies and with similar angular resolution in multiple observing epochs, our ability the investigate flux density variability is limited. However, the targets can be investigated by comparing the high-resolution VLBI flux densities with single-dish and low-resolution radio interferometric data. It should be noted that the estimated total flux density values themselves may be affected by long-term variability since the measurements at different frequencies used to determine the spectral shapes and spectral indices were not simultaneous. If $S_{\mathrm{1.7}}/S_{\mathrm{VLA,1.7}} > 1.1$, we consider the source variable. Only two of the sources (J1614$+$4640 and J2239$+$0030) fulfil this criterion, although the large uncertainties make the possible variabilities indefinite. The possible explanation for ratios below 0.9 is that a significant fraction of the total radio emission originates from an extended region resolved out with VLBI. Nevertheless, variability cannot be ruled out in these cases, just the data available at present are not sufficient to claim it.

\citet{2022ApJS..260...49K} found about half of their 13 targets in a $4 \lesssim z \lesssim 4.5$ radio quasar sample to be variable, while in a larger $z > 4.5$ sample containing the then-known 30 VLBI-detected source, \citet{2016MNRAS.463.3260C} concluded that only about $20\%$ of the targets are significantly variable. In recent studies, \citet{2023MNRAS.518...39W,2023MNRAS.525.6064W} investigated RQQs and RLQs at $z \lesssim 0.5$ and found that core-dominated RQQs with flat/inverted spectrum have more pronounced variability, while steep-spectrum jet-dominated RQQs show negligible variability. Our findings, albeit limited by the available data, do not contradict with the findings cited above.

\subsection{The non-detection of J0306+1853}

J0306$+$1853 remained undetected in our EVN--e-MERLIN observation. Earlier, \citet{2021AA...655A..95S} observed this source with the VLA at 1.5 and 5~GHz. At both frequencies, they found the source unresolved with point-like structure on arcsecond scale. The derived spectral index ($-0.87\pm0.06$, see also our Table~\ref{tab:sedparams}) indicates a steep spectrum and thus an extended structure. However, the VLA detection raised the possibility of the presence of a sub-arcsecond compact jet structure. The radio loudness $R_{4400\AA} = 0.7$ \citep{2021AA...655A..95S} indicates that the source is radio-quiet. The sensitivity of the EVN observation allowed us to reach $113~\mu$Jy\,beam$^{-1}$ ($5~\sigma$) detection limit, which would have been sufficient to detect a milliarcsecond-scale compact source with the 1.5-GHz VLA flux density of 252~$\mu$Jy with a signal-to-noise ratio of $\sim 10$. Based on our non-detection, the radio emission of J0306$+$1853 is apparently resolved out on milliarcsecond to $\sim0\farcs1$ angular scales. It seems less likely but still possible that J0306$+$1853 is variable.

\subsection{The radio loudness--brightness temperature relation at high redshift}\label{subsec:r-tb}
 
There are two commonly used definitions of the radio-loudness parameter. \citet{1989AJ.....98.1195K,2016ApJ...831..168K} defined $R_{4400\AA}$ as the fraction of the rest-frame 5-GHz radio and the optical power at 4400~\AA. On the other hand, \citet{1980ApJ...238..435S} defined $R_{2500\AA}$ based on the 2500~\AA\ optical power. To investigate the relation between radio loudness and other physical parameters, focusing on high-redshift quasars, we collected all the available $z > 5$ sources from the literature, altogether 48 objects, which have either any radio-loudness index ($R_{4400\AA}$ or $R_{2500\AA}$) published, or L-band ($1.4-1.7$~GHz) VLBI observation available, or both. SMBH masses were also collected from the literature where available, for about one third of the sample. We found that the mean $M_{\mathrm{SMBH}}$ at $z > 5$ is $4.4 \times 10^9$~M$_\odot$ All the collected sources are listed in Table~\ref{tab:references} with their radio-loudness indices, brightness temperature, SMBH mass, L-band flux densities and monochromatic powers, along with the respective references. The choice of the L band is motivated by the fact that this is the most commonly used frequency band for $z > 5$ quasar VLBI observations, and this way we can also include our nine newly-detected sources in the sample.

\begin{figure*}[!ht]
    \centering
    \includegraphics[width=0.49\textwidth]{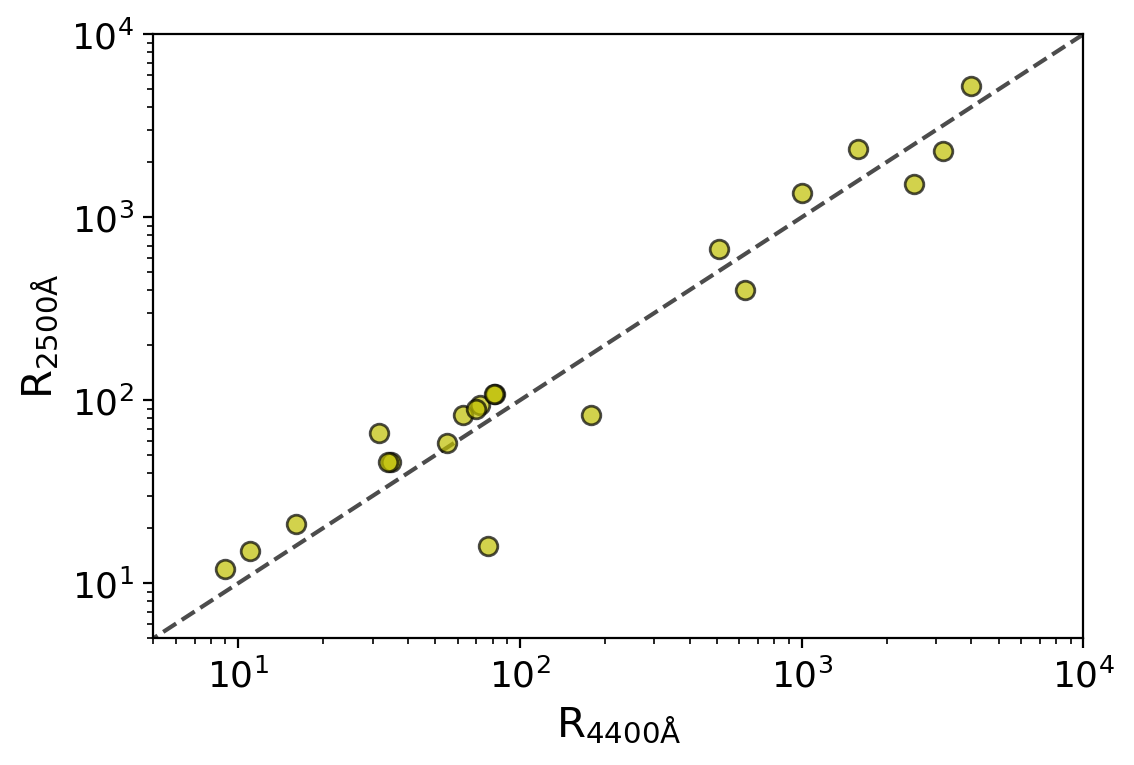}
    \includegraphics[width=0.49\textwidth]{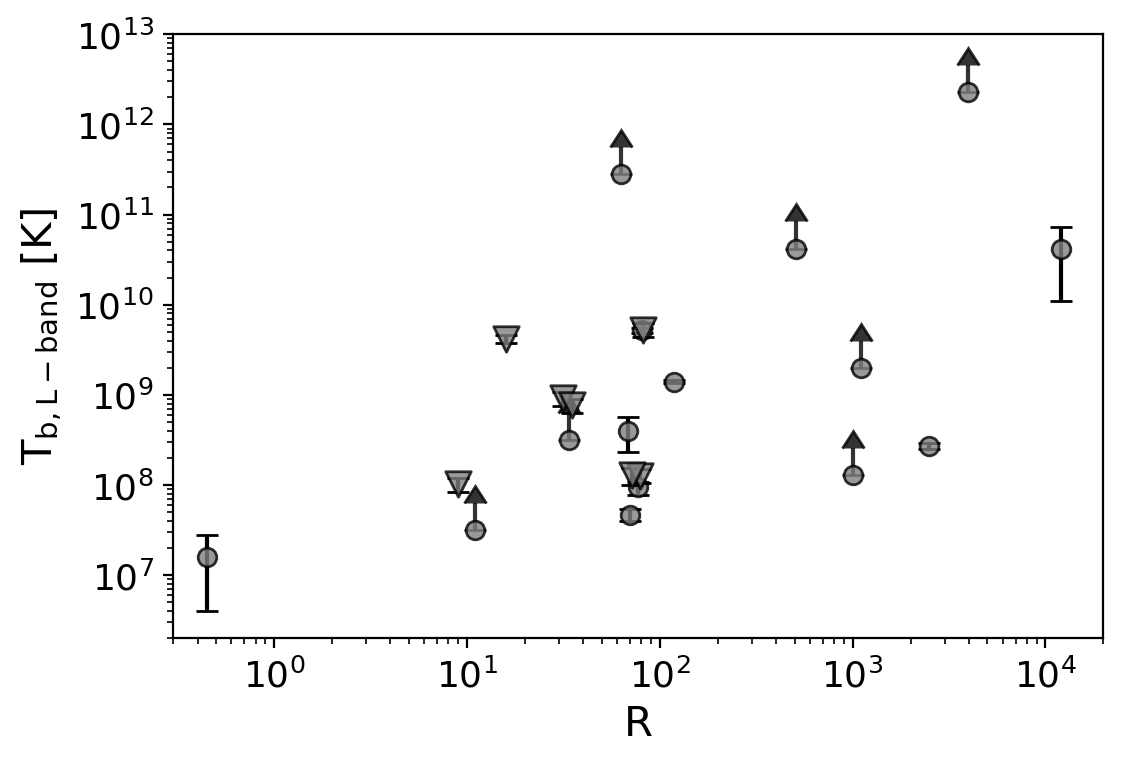}
    \includegraphics[width=0.49\textwidth]{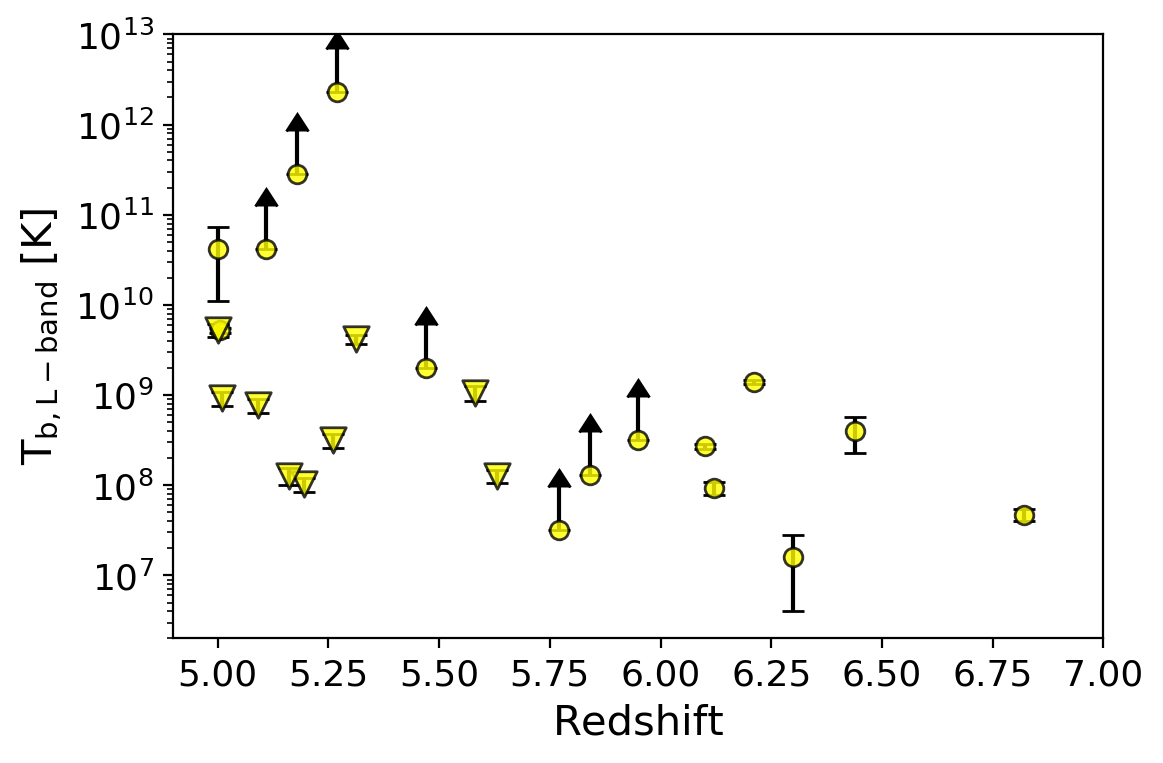}
    \includegraphics[width=0.49\textwidth]{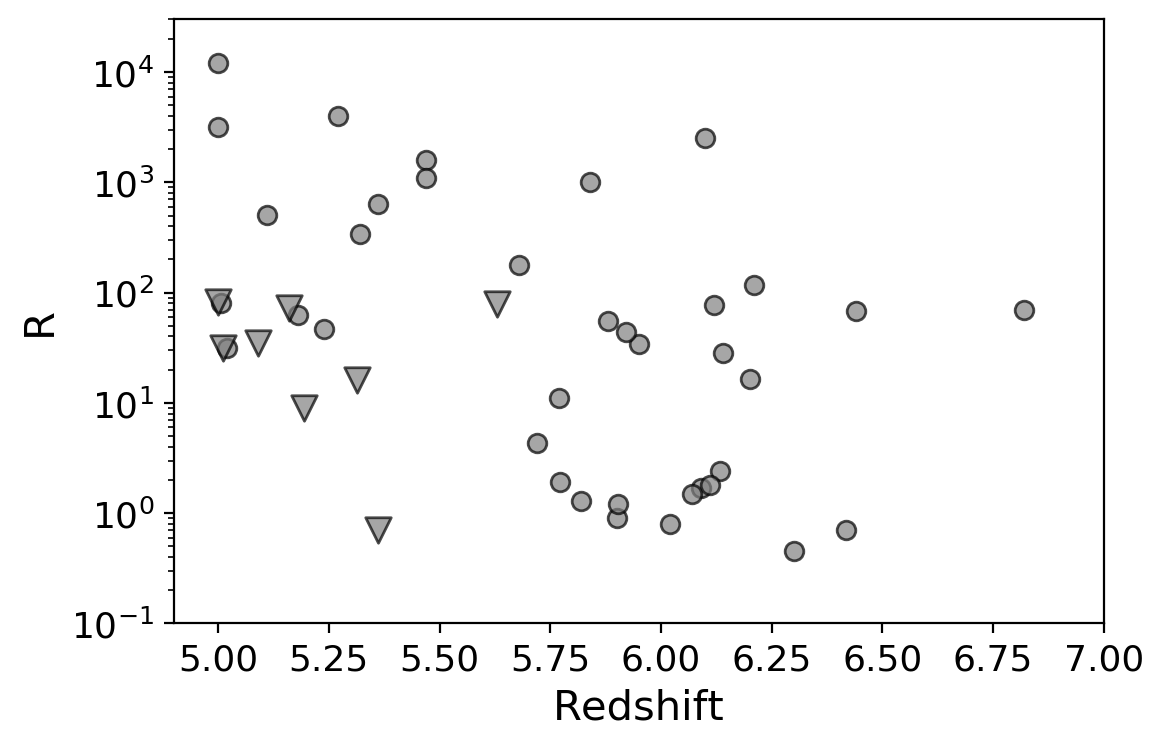}
    \caption{Radio-loudness and brightness temperature relation of high redshift radio sources.
    \textit{Top left:} The $R_{\mathrm{2500\,\AA}}$ index as a function of $R_{\mathrm{4400\,\AA}}$ for sources where both values are available in the literature. The dashed line represents $R_{\mathrm{2500\,\AA}} = R_{\mathrm{4400\,\AA}}$.
    \textit{Top right:} The redshift-corrected L-band brightness temperatures as a function of radio loudness. The sources presented in this paper are marked with triangles.  
    \textit{Bottom left:} The redshift-corrected L-band brightness temperature measurements for all $z>5$ AGN published to date, including those presented in this paper (triangles), as a function of redshift.
    \textit{Bottom right:} The $R_{\mathrm{4400\,\AA}} = L_{\mathrm{5\,GHz}}/L_{\mathrm{4400\,\AA}}$ and $R_{\mathrm{2500\,\AA}} = L_{\mathrm{5\,GHz}}/L_{\mathrm{2500\,\AA}}$ (where $R_{\mathrm{4400\,\AA}}$ is not available) radio-loudness indices of $z>5$ sources as a function of redshift.}
    \label{fig:rl_vs_tb}
\end{figure*}

For the VLBI observations, we took data and references from Table~1 of \citet{2022ApJS..260...49K} as a starting point, and collected the derived redshift-corrected brightness temperatures. We extended this sample with the few sources that have been published since then: VIK J2318$-$3113 \citep{2022AA...662L...2Z}, J0141$-$5427 \citep{2023PASA...40....4G}, and J1702$+$1301 \citep{2024A&A...685A.111L}. There are currently 24 sources known with L-band VLBI observations, including the nine presented in this paper. Out of these 24 sources, seven have only lower limit to $T_{\mathrm b}$. Each source has a single epoch of observation. The bottom left panel in Fig.~\ref{fig:rl_vs_tb} shows the L-band brightness temperatures as a function of redshift. The median $T_{\mathrm b}$ is $5.9 \times 10^8$~K. The measured $T_{\mathrm b}$ values show a generally decreasing trend with increasing redshift. Also, the jet emission of only a small fraction of these sources appears Doppler-boosted ($T_{\mathrm b} \gtrsim 5 \times 10^{10}$~K) at L-band. 

We found 43 sources with published $R_{4400\AA}$ and 23 sources with $R_{2500\AA}$ values, with 21 overlapping objects. In the cases where multiple $R$ values are published for a given source, we considered the value published most recently. Because of variability, the $R$ value may depend on the epochs when the luminosities were measured. It is usually not feasible to measure the radio and optical luminosities at the same time. Therefore strong variability of certain sources can influence the $R$ indices when there are optical magnitudes or radio flux density measurements available at different epochs. Another uncertainty of $R$ is caused by the choice of spectral indices assumed when performing K-correction, i.e. calculating the rest-frame 5-GHz or 4400\,$\AA$ luminosities. Considering all these caveats, it is not surprising to find occasionally very different $R$ values in the literature for many sources. However, the plot showing $R_{2500\AA}$ as a function of $R_{4400\AA}$ values (top left panel of Fig.~\ref{fig:rl_vs_tb}) for the $z>5$ sources where both values are available indicates that in most cases the values of the two radio-loudness indices are not markedly different. This justifies their use together when analysing the general properties of the sample.

The top right panel of Fig.~\ref{fig:rl_vs_tb} displays the distribution of $T_{\mathrm b}$ as a function of radio-loudness parameters $R_{4400\AA}$ and $R_{2500\AA}$. Based on this plot, a general trend of increasing brightness temperature with increasing radio loudness can be seen. It appears that, despite the uncertainties described above, the sources with higher $R$ index have higher $T_{\mathrm b}$ in L band. Since the radio loudness is proportional to the total flux density and the brightness temperature to the core flux density (Eq.~\ref{eq:tb}), the existence of such a relation is not surprising if the radio emission is confined to the compact AGN core region. Populating the parameter space where the coverage is more incomplete (i.e. $R < 10$ and $R > 100$), this relation could be constrained further. The high-redshift regime of the $R-z$ diagram started to be built up in the recent years \citep{2019AA...629A..68B,2021ApJ...909...80B}, with more and more distant extragalactic radio sources identified. The bottom right panel of Fig.~\ref{fig:rl_vs_tb} shows the $R_{4400\AA}$ and $R_{2500\AA}$ (where there was no available $R_{4400\AA}$ value) indices of AGN at $z > 5$ as a function of redshift. The median values of the collected $R_{4400\AA}$ and $R_{2500\AA}$ indices are are 35 and 90, respectively.

\subsection{The radio power distribution at high redshift}

The derived core and total monochromatic radio powers of our sources at $1.7$~GHz are in the range of $\sim 10^{25-27}$~W\,Hz$^{-1}$. These radio powers are typical of Fanaroff--Riley type II \citep[FRII,][]{1974MNRAS.167P..31F} radio galaxies and quasars, spanning from the low-end to the the high-power regime \citep[e.g.][]{2001ApJ...552..508G, 2004Ap&SS.293....1G,2021AA...655A..95S}. The phenomenological
division line between FRI and FRII sources is at $P_\mathrm{total} \approx 10^{25}$~W\,Hz$^{-1}$ \citep{1974MNRAS.167P..31F}, but it is based on low-frequency observations. It appears that while these sources can be classified as radio-quiet, they in fact have radio powers typical of a 3C radio galaxy, i.e. $\gtrsim 10^{25}$~W\,Hz$^{-1}$ \citep[e.g.][]{2001ApJ...552..508G, 2004Ap&SS.293....1G,2021AA...655A..95S}.

We checked the sample of $z>5$ sources collected from the literature for published L-band total and core flux densities and monochromatic radio powers. We used the flux densities to calculate the total ($P_\mathrm{total}$) and core ($P_\mathrm{core}$) K-corrected monochromatic radio powers (see Eq. \ref{eq:power}) in cases where these values were not available. We assumed $\alpha = 0$ for calculating all $P_{\mathrm{total}}$ values, and those $P_{\mathrm{core}}$ values where $\alpha$ was not published. The collected parameters are listed in Table~\ref{tab:references} along with their references.

\begin{figure*}[!ht]
    \centering
    \includegraphics[width=0.506\textwidth]{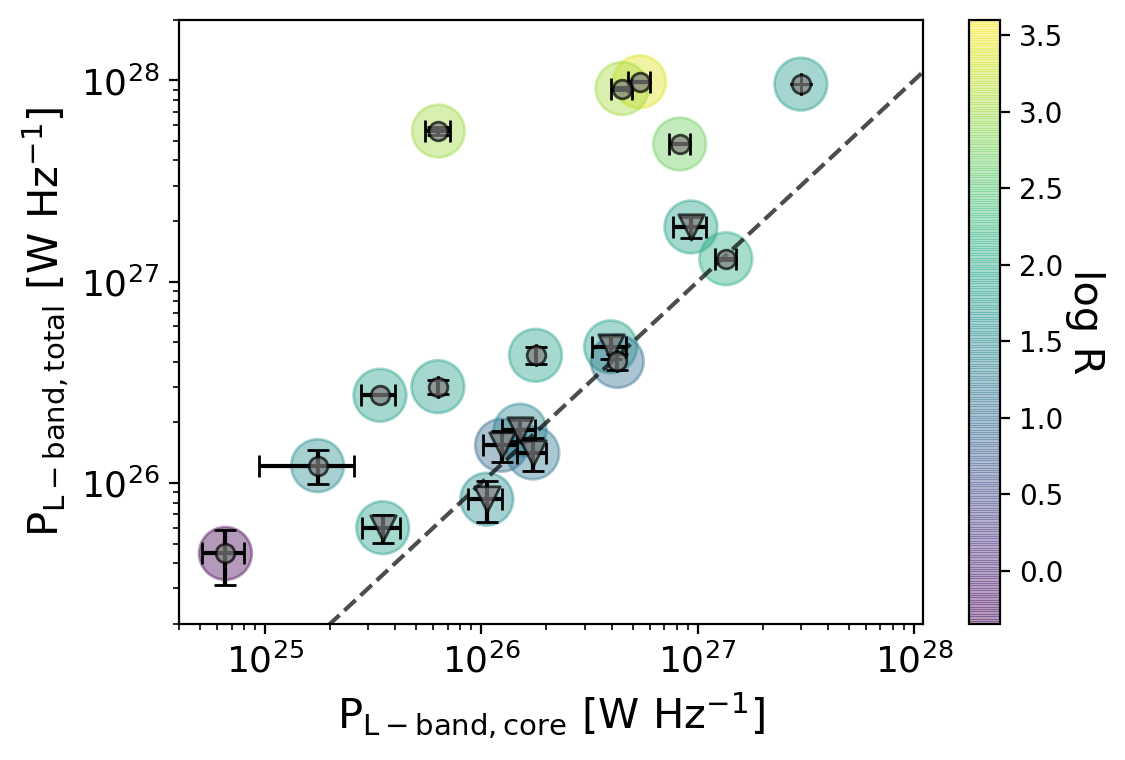}
    \includegraphics[width=0.49\textwidth]{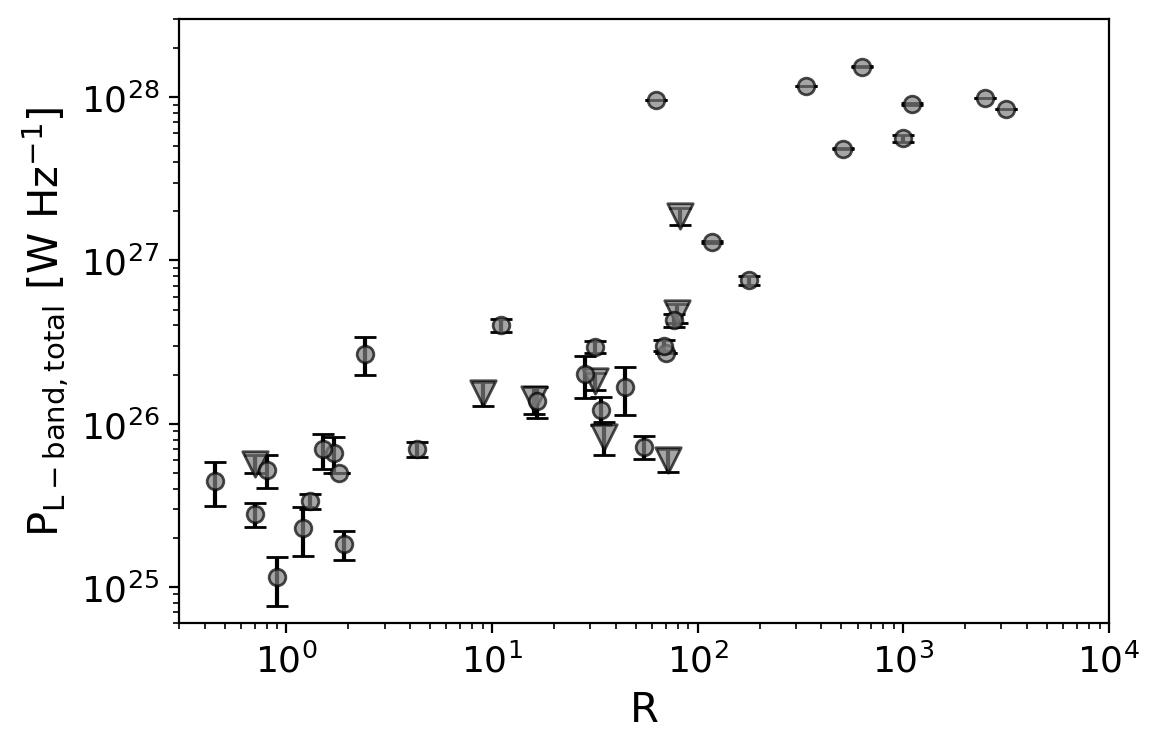}
    \includegraphics[width=0.49\textwidth]{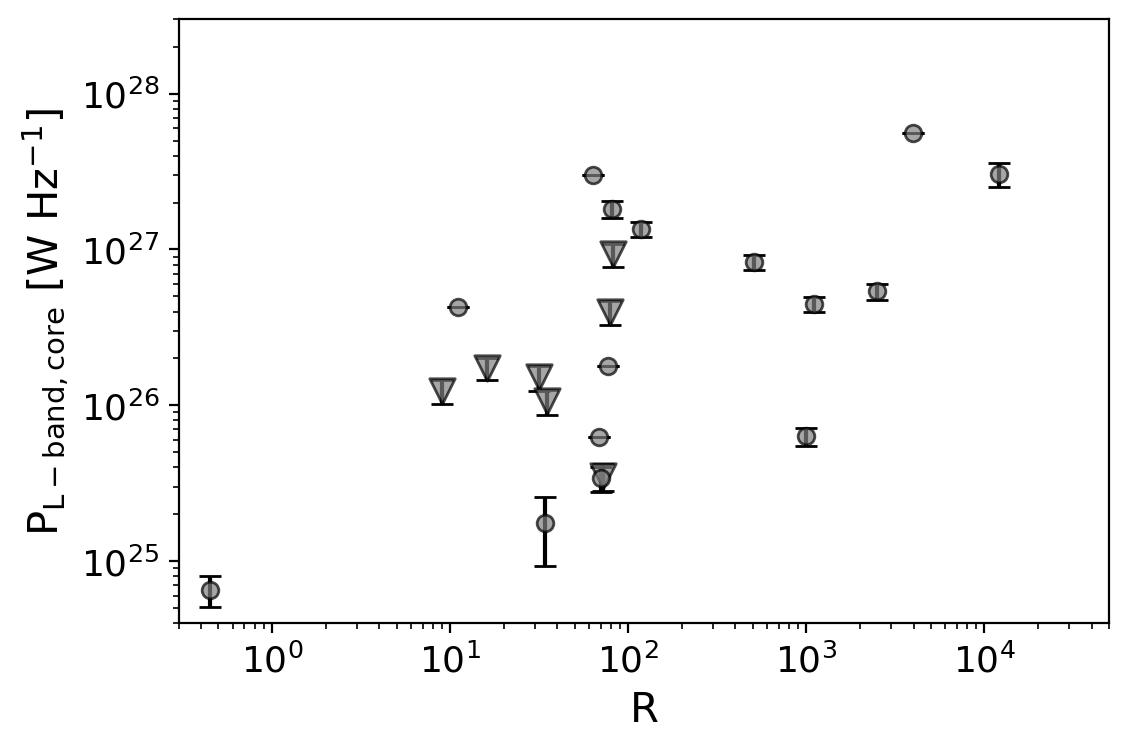}
    \includegraphics[width=0.49\textwidth]{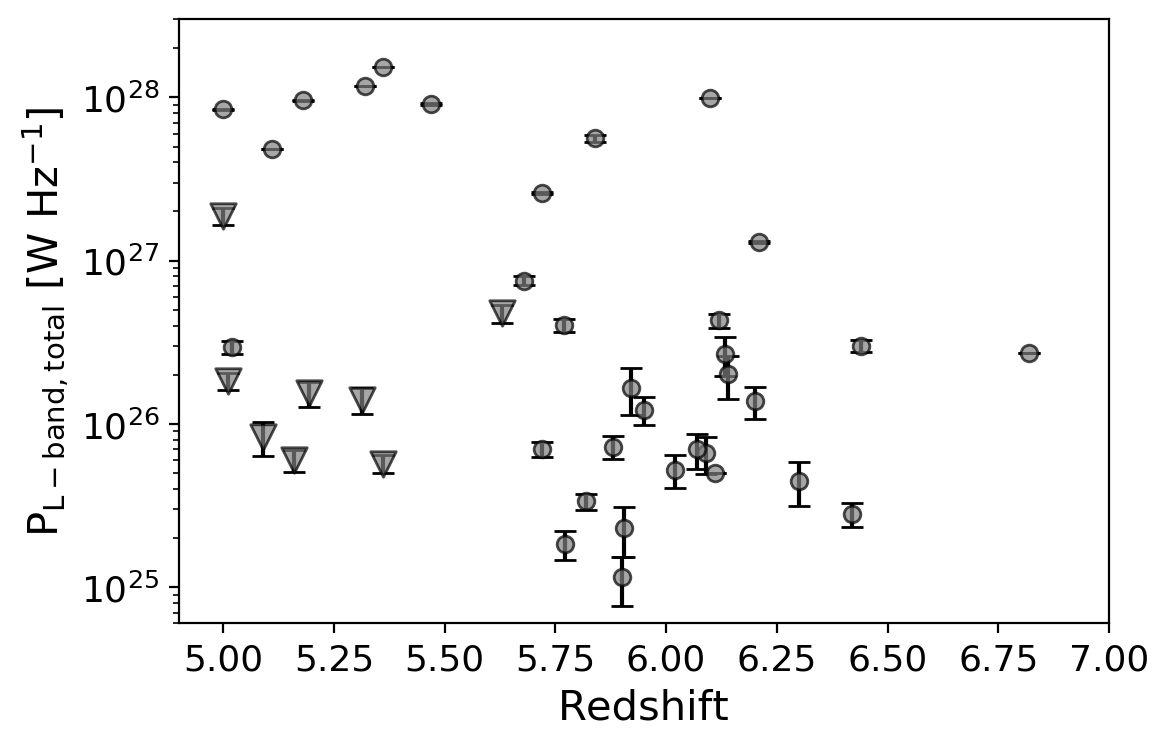}
    \includegraphics[width=0.49\textwidth]{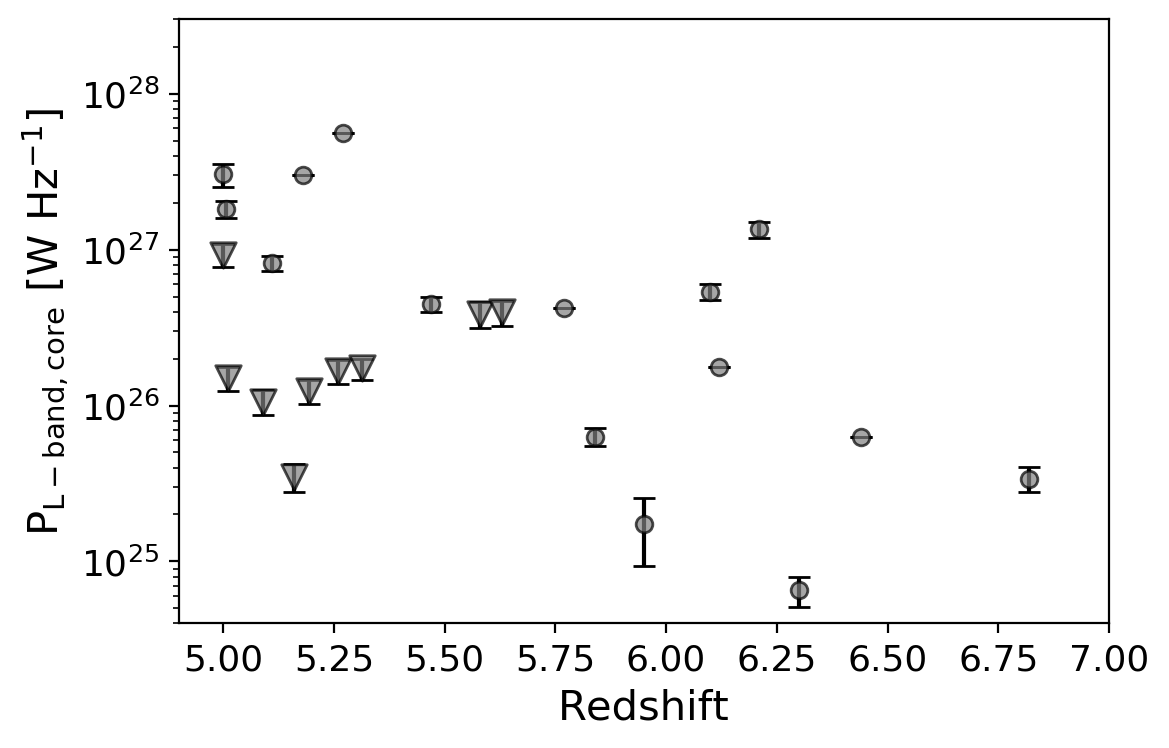}
    \caption{L-band total and core monochromatic radio powers of high redshift radio sources.
    The sources presented in this paper are marked with triangles. \textit{Top:} The comparison of L-band core monochromatic radio power and the L-band total monochromatic radio power for $z>5$ sources where both values are available. The dashed line represents the $P_{\mathrm{total}} = P_{\mathrm{core}}$. The colours indicate the source radio loudness. \textit{Middle:} The total \textit{(left)} and core \textit{(right)} monochromatic radio powers of $z>5$ sources as a function of radio loudness. \textit{Bottom:} The total \textit{(left)} and core \textit{(right)} monochromatic radio powers as a function of redshift.}
    \label{fig:rl_vs_p} 
\end{figure*}

The top panel of Fig.~\ref{fig:rl_vs_p} shows the comparison of $P_{\mathrm{total}}$ and $P_{\mathrm{core}}$ for $z > 5$ sources, where both values were available. The dashed line indicates $P_\mathrm{total} = P_\mathrm{core}$. Sources lying above this line have more extended emission, since the total power is higher than the core power. The sources become more core-dominated and more likely to be beamed as the distance from the line decreases. The sources scattering close to the $P_\mathrm{total} = P_\mathrm{core}$ line or slightly below, like some of our sources, can also be affected by variability. The colouring in the top panel of Fig.~\ref{fig:rl_vs_p} reveals that in general the higher the radio loudness the less core-dominated the source appears. This implies that the radio emission responsible for the high radio loudness tends to originate from the kiloparsec-scale radio structure rather than the compact parsec-scale region.

We found that our sources, although appearing core-dominated, do not show evidence for relativistically beamed radio jets based on the derived brightness temperatures (Table~\ref{tab:physparams}). One reason could be that the $T_{\mathrm b}$ values are underestimated, by slightly overestimating the modelled size of the VLBI core component. However, this cannot be responsible for the nearly $2$ orders of magnitude difference from the equipartition brightness temperature. A possible physical explanation is that the radio emission region of these sources is intrinsically small, confined to sub-kiloparsec scales. Gigahertz-peaked spectrum (GPS) radio sources and compact symmetric objects (CSOs) \citep[e.g.][]{1994ApJ...432L..87W,2012ApJ...760...77A,2021A&ARv..29....3O} are known to be young jetted AGN with compact radio structure, without beamed radio jets and diffuse extended emission. In fact, a few known CSOs can be found among high-redshift sources \citep[e.g.][]{2008AA...484L..39F,2008AJ....136..344M,2022MNRAS.511.4572A,2022ApJS..260...49K}. The GPS and CSO nature of the sources could be supported by their peaked spectrum \citep[e.g.][]{2016MNRAS.459..820T}. In our sample, only four of the sources have peaked spectra (Fig.~\ref{fig:sed}), but except for J2239$+$0030, all the others have poor low-frequency coverage in their flux density measurements, retaining the possibility that their spectrum is actually peaked. 

The middle panels of Fig.~\ref{fig:rl_vs_p} display the relation between the radio-loudness indices and the derived radio powers, similar to the $T_{\mathrm{b}} - R$ plot in Sect.~\ref{subsec:r-tb}. Both the $P_{\mathrm{total}}$ and $P_{\mathrm{core}}$ values show a steadily increasing trend. Since $R$ is proportional to the $5$-GHz total radio power, this trend is not surprising for $P_{\mathrm{total}}$. The bottom panels of Fig.~\ref{fig:rl_vs_p} show the $P_{\mathrm{total}}$ and $P_{\mathrm{core}}$ values for AGN at $z > 5$ as a function of redshift. Median values for both the collected $P_{\mathrm{total}}$ and $P_{\mathrm{core}}$ are around $2\times10^{26}$~W~Hz$^{-1}$.

\section{Conclusion} \label{conclusion}

In this paper, we present the results of 1.7-GHz VLBI observation of ten high-redshift ($5< z < 6$) radio quasars using the EVN and e-MERLIN arrays. The sources have different radio-loudness indices spanning a wide range of $\sim 10^{-1}-10^2$. Out of the ten sources, nine have been successfully detected. They all show a single faint but compact milliarcsecond-scale radio core component. The 90\% detection rate for this faint sample is impressive and demonstrates the effectiveness of the observing strategy and the sensitivity of the VLBI network used. This opens the way for future VLBI studies of many more similar $z > 5$ sources. Further improvements in VLBI sensitivity expected in the future from e.g. the first phase of the Square Kilometre Array \citep{2015aska.confE.143P} or the Five-hundred-meter Aperture Spherical Radio Telescope core array \citep{2024AstTI...1...84J} are promising for advancing this work.

The sample presented here increases the number of known VLBI-observed $z > 5$ radio AGN by about $40\%$. These high-resolution interferometric observations, combined with additional low-resolution radio data obtained from the literature, helped derive properties such as the origin of the radio emission, compactness, flux density variability, and spectral indices. The brightness temperatures obtained confirm the AGN origin of the radio emission and suggest that the emission in none of these sources is Doppler-enhanced. The derived monochromatic powers are comparable to those of the luminous FRII radio galaxies and quasars. As J0306$+$1853 remained undetected at milliarcsecond-scale, we have not been able to reveal the compact jet expected by \citet{2021AA...655A..95S}. We found that five of the sources have \textit{Gaia} optical astrometric positions. One of them is J0306$+$1853 with no VLBI detection, but the optical positions of the other four AGN were found to be consistent with the VLBI radio core positions within the uncertainties. 

Notably, we did not find crucial differences in the morphology and the derived physical parameters between our detected three RQQ and six RLQ sources. These sources appear core-dominated with high radio power, despite being in the transition region of RQQs to RLQs. With no compelling evidence for Doppler-boosted jet emission, it is possible that these are young jetted AGN whose radio emission is confined to a small region around the central SMBH.

By examining this sample, supplemented with other $z > 5$ sources detected with VLBI at L-band from the literature, we found that the core brightness temperatures and monochromatic radio powers tend to increase with increasing radio loudness. However, for the $R-T_{\mathrm b}$ relation, the radio-quiet and the extreme radio-loud regimes are still undersampled. High-resolution VLBI observations of many more radio-emitting AGN at $z>5$ would be desirable, even though it is challenging because the spectral turnover is shifting to lower frequencies as $z$ increases. Sensitive higher-frequency and therefore higher-resolution VLBI observations of the investigated nine detected sources would provide tighter constraints on the brightness temperature, and information on the spectral properties of the cores.

\begin{acknowledgements}
The EVN is a joint facility of independent European, African, Asian and North American radio astronomy institutes. Scientific results from data presented in this publication are derived from the following EVN project code: EG102. The e-MERLIN is a National Facility operated by the University of Manchester at Jodrell Bank Observatory on behalf of STFC. 
This work presents results from the European Space Agency (ESA) space mission \textit{Gaia}. \textit{Gaia} data are being processed by the \textit{Gaia} Data Processing and Analysis Consortium (DPAC). Funding for the DPAC is provided by national institutions, in particular the institutions participating in the \textit{Gaia} MultiLateral Agreement (MLA). The \textit{Gaia} mission website is \url{https://www.cosmos.esa.int/gaia}. The \textit{Gaia} archive website is \url{https://archives.esac.esa.int/gaia}.
This research has made use of the NASA/IPAC Extragalactic Database (NED) which is operated by the Jet Propulsion Laboratory, California Institute of Technology, under contract with the National Aeronautics and Space Administration.
M.K. thanks for the support from the \'UNKP-23-3 New National Excellence Program of the Ministry for Culture and Innovation from the source of the National Research, Development and Innovation Fund. We thank the Hungarian National Research, Development and Innovation Office (NKFIH, grant no. OTKA K134213) for support. This work was also supported by HUN-REN and the NKFIH excellence grant TKP2021-NKTA-64. 
T.A. thanks for the support from the National SKA Program of China (2022SKA0120102).
\end{acknowledgements}

\bibpunct{(}{)}{;}{a}{}{,}
\bibliographystyle{aa}
\bibliography{main}

\begin{appendix}
\section{Radio loudness and brightness temperatures}
\begin{sidewaystable*}[!ht]
    \centering
    \scriptsize
    \caption{Values and references for the collected radio-loudness indices, L-band brightness temperatures, core flux densities, monochromatic powers, and black hole masses for the $z>5$ radio AGN listed with increasing redshift.}
    \begin{tabular}{clrcrcrcrcrcrcrcrcr}
\hline\hline
Redshift & Source ID & $R_{\mathrm{4400\,\AA}}$ & Ref. & $R_{\mathrm{2500\,\AA}}$ & Ref. & $T_{\mathrm{b,L-band}}$ & Ref. & $S_{\mathrm{total,L-band}}$ & Ref. & $S_{\mathrm{core,L-band}}$ & Ref. & $P_{\mathrm{total,L-band}}$ & Ref. &  $P_{\mathrm{core,L-band}}$ & Ref. & $M_{\mathrm{SMBH}}$ & Ref. \\
 & & & & & & [$10^8$ K] & & [mJy] & & [mJy] & & [$10^{26}$ W~Hz$^{-1}$] & & [$10^26$ W~Hz$^{-1}$] & & $10^8 M_\odot$ & \\
        \hline
5.00 & J2344$+$1653 & 82 & 32 & 108 & 32 & 53.0 (8.9) & 0  &   &  & 7.0 (0.9) & 0 & 18.7 (2.3) & 0 & 9.3 (1.6) & 0 &  &  \\
5.00 & J0141$-$5427 &  &  & 12000 & 5 & 420 (310) & 18 & 160 & 5 & 80.3 (8.4) & 18 & 419 & 0 & 30.4 (5.3) & 0 & 8 & 5 \\
5.00 & GB6J162956$+$095959 & 3162 & 8 & 2300 & 5 &   &  & 32.3 (0.1) & 35 &   &  & 84.4 (0.3) & 0 &   &  &  &  \\
5.01 & J1146$+$4037 & 81 & 4 & 108 & 4 & 52.1 (2.9) & 14  &   &  & 15.5 (0.8) & 14  &   &  & 18.3 (2.3) & 14  &  &  \\
5.01 & J1034$+$2033 & 32 & 8 & 66 & 4 & 9.2 (1.6) & 0 & 3.9 (0.1) & 35 & 3.5 (0.5) & 0 & 1.8 (0.2) & 0 & 1.5 (0.3) & 0 &  &  \\
5.02 & J1013$+$3518 & 32 & 8 &  &  &   &  & 1.1 (0.1) & 35 &   &  & 3.0 (0.3) & 0 &   &  &  &  \\
5.09 & J2239$+$0030 & 35 & 32 & 46 & 32 & 7.7 (1.4) & 0 & 1.4 (0.1) & 35 & 1.4 (0.2) & 0 & 0.8 (0.2) & 0 & 1.1 (0.2) & 0 &  &  \\
5.11 & J0913$+$5919 & 509 & 4 & 674 & 4 & 420  & 38 & 17.6 (0.1) & 35 & 19.4 (0.1) & 38 & 48.3 (0.3) & 0 & 8.3 (0.9) & 0 &  &  \\
5.16 & J2245$+$0024 & 72 & 4 & 94 & 4 & 1.3 (0.3) & 0 & 0.9 (0.1) & 35 & 0.8 (0.1) & 0 & 0.6 (0.1) & 0 & 0.4 (0.1) & 0 &  &  \\
5.18 & J0131$-$0321 & 63 & 4 & 83 & 4 & 2800  & 16 & 33.7 (0.1) & 35 & 64.4 (0.3) & 16 & 95.8 (0.3) & 0 & 30 & 16 & 110 & 19 \\
5.19 & J0741$+$2521 & 9 & 4 & 12 & 4 & 1.0 (0.2) & 0 & 2.9 (0.1) & 35 & 2.7 (0.4) & 0 & 1.5 (0.3) & 0 & 1.2 (0.2) & 0 &  &  \\
5.24 & J2329$+$3003 &  &  & 47 & 26 &   &  &   &  &   &  &   &  &   &  &  &  \\
5.26 & J0747$+$1153 &  &  &  &  & 3.1 (0.6) & 0 &  &  & 1.8 (0.2) & 0 &  &  & 1.7 (0.3) & 0 &  &  \\
5.27 & J1026$+$2542 & 3981 & 8 & 5200 & 27 & 23000  & 15  & 239.4 (0.1) & 35 & 180.4 (0.5) & 15  & 708.9 (0.3) & 0 & 56  & 15  &  &  \\
5.31 & J1614$+$4640 & 16 & 4 & 21 & 4 & 41.7 (4.5) & 0 & 2.2 (0.1) & 35 & 3.6 (0.4) & 0 & 1.4 (0.3) & 0 & 1.7 (0.3) & 0 &  &  \\
5.32 & PSO J191$+$86 & 337 & 7 &  &  &   &  & 74.2 (2.3) & 10 &   &  & 117  & 7 &   &  & 24 & 7 \\
5.36 & J0306$+$1853 & 0.7 & 28 &  &  &   &  & 0.3  & 28 &   &  & 0.6 (0.1) & 0 &   &  & 110 & 36 \\
5.36 & GB6J164856$+$460341 & 631 & 8 & 400 & 5 &   &  & 49.9 (0.1) & 35 &   &  & 153.6 (0.3) & 0 &   &  &  &  \\
5.47 & J1702$+$1301 & 1100 & 32 &  &  & 20  & 24 & 28.3 (0.4) & 24 & 8.4 (0.1) & 24 & 91.2 (1.3) & 0 & 4.5 (0.5) & 0 &  &  \\
5.47 & J0906$+$6930 & 1585 & 8 & 2373 & 26 &   &  &   &  &   &  &   &  &   &  &  &  \\
5.58 & J0616$-$1338 &  &  &  &  & 10.6 (2.0) & 0 &   &  & 1.9 (0.3) & 0 &   &  & 3.9 (0.7) & 0 &  &  \\
5.63 & J0901$+$1615 & 79.4 & 8 &  &  & 1.3 (0.2) & 0 & 3.9 (0.1) & 35 & 3.1 (0.4) & 0 & 4.8 (0.6) & 0 & 3.9 (0.7) & 0 &  &  \\
5.68 & PJ055$-$00 & 178 & 2 & 83 & 26 &   &  & 2.1 (0.1) & 2 &   &  & 7.5 (0.5) & 0 &   &  &  &  \\
5.72 & J1530$+$1049 &  &  &  &  &   &  & 7.3 (0.1) & 35 & 1.7  & 17 & 25.9 (0.4) & 0 &  0.9  & 0  &  &  \\
5.72 & J0203$+$0012 & 4.3 & 2 &  &  &   &  & 0.20 (0.02) & 2 &   &  & 0.7 (0.1) & 0 &   &  &  &  \\
5.77 & 0836$+$0054 & 11 & 4 & 15 & 4 & 0.3 & 12 & 1.1 (0.1) & 35 & 1.1 & 12 & 4.0 (0.4) & 0 & 4.3 & 12 & 83 & 30 \\
5.77 & J0927$+$2001 & 1.9 & 2 &  &  &   &  & 0.05 (0.01) & 2 &   &  & 0.18 (0.04) & 0 &   &  & 49 & 30 \\
5.82 & J0002$+$2550 & 1.3 & 2 &  &  &   &  & 0.09 (0.01) & 2 &   &  & 0.34 (0.04) & 0 &   &  & 48 & 30 \\
5.84 & P352$-$15 & 1000 & 25 & 1352 & 26 & 1.3  & 25 & 14.9 (0.7) & 10 & 1.2 (0.1) & 25 & 56.0 (2.6) &  & 0.6 (0.1) & 0 &  &  \\
5.88 & J2242$+$0334 & 55 & 22 & 58 & 26 &   &  & 0.19 (0.03) & 22 &   &  & 0.7 (0.1) & 0 &   &  &  &  \\
5.90 & J1335$+$3533 & 0.9 & 2 &  &  &   &  & 0.03 (0.01) & 2 &   &  & 0.12 (0.04) & 0 &   &  & 29 & 30 \\
5.90 & J1411$+$1217 & 1.2 & 2 &  &  &   &  & 0.06 (0.02) & 2 &   &  & 0.2 (0.1) & 0 &   &  & 5 & 11 \\
5.92 & J2053$+$0047 & 44 & 2 &  &  &   &  & 0.4 (0.1) & 2 &   &  & 1.7 (0.5) & 0 &   &  &  &  \\
5.95 & J2228$+$0110 & 34 & 4 & 46 & 4 & 3.1 & 9 & 0.3 (0.1) & 2 & 0.3 (0.1) & 9 & 1.2 (0.2) & 0 & 0.2 (0.1) & 0 &  &  \\
6.02 & J0818$+$1722 & 0.8 & 21 &  &  &   &  & 0.13 (0.03) & 21 &   &  & 0.5 (0.1) & 0 &   &  &  &  \\
6.07 & J1034$-$1425 & 1.5 & 21 &  &  &   &  & 0.17 (0.04) & 21 &   &  & 0.7 (0.2) & 0 &   &  &  &  \\
6.09 & J1602$+$4228 & 1.7 & 21 &  &  &   &  & 0.16 (0.04) & 21 &   &  & 0.7 (0.2) & 0 &   &  & 24 & 30 \\
6.10 & PSO J0309$+$27 & 2500 & 6 & 1521 & 26 & 2.7 (0.2) & 31 & 23.7 (0.1)  & 10 & 8.0 (0.1) & 31 & 98.5  &  & 5.4 (0.6) & 0 &  &  \\
6.11 & J1558$-$0724 & 1.8 & 21 &  &  &   &  & 0.12 (0.04) & 21 &   &  & 0.5  & 0 &   &  &  &  \\
6.12 & J1427 $+$3312 & 77 & 21 & 16 & 4 & 0.9 (0.2) & 13 & 1.0 (0.1) & 35 & 1.5  & 13 & 4.3 (0.4) & 0 & 1.8  & 13 & 11 & 30 \\
6.13 & J1319$+$0950 & 2.4 & 2 &  &  &   &  & 0.6 (0.2) & 22 &   &  & 2.7 (0.7) & 0 &   &  & 21 & 34 \\
6.14 & J1609$+$3041 & 28.3 & 2 &  &  &   &  & 0.5 (0.1) & 2 &   &  & 2.0 (0.6) & 0 &   &  & 43 & 30 \\
6.20 & J0227$-$0605 & 16.5 & 22 &  &  &   &  & 0.3 (0.1) & 22 &   &  & 1.4 (0.3) & 0 &   &  & 2 & 30 \\
6.21 & J1429$+$5447 & 118 & 21 &  &  & 14.0 (0.6) & 37 & 3.00 (0.04) & 21 & 3.3 (0.1) & 37 & 12.9 (0.2) & 0 & 13.5 (1.5) & 37 & 18 & 30 \\
6.30 & J0100$+$2802 & 0.5 & 28 &  &  & 0.2 (0.1) & 23 & 0.10 (0.03) & 21 & 0.09 (0.02) & 23 & 0.5 (0.1)  & 0 & 0.07 (0.01) & 0 & 120 & 36 \\
6.42 & J1148$+$5251 & 0.7 & 2 &  &  &   &  & 0.06 (0.01) & 2 &   &  & 0.3 (0.1) & 0 &   &  & 87 & 30 \\
6.44 & VIK J2318$-$3113 & 69 & 21 &  &  & 4.0 (1.7) & 33 & 0.6 (0.1) & 21 & 0.6 (0.1) & 33 & 3.0 (0.2) & 0 & 0.63  & 33 & 6 & 20 \\
6.82 & P172$+$18 & 70 & 26 & 90 & 26 & 0.5 (0.1) & 26 & 0.5 & 3 & 0.4 (0.1) & 26 & 2.7 & 0 & 0.3 (0.1) & 0 &  &  \\
        \hline      
\end{tabular}
    \newline
Notes. Col.~1 -- source redshift; Col.~2 -- source designation; Col.~3-4 -- $R_{\mathrm{4400\,\AA}} = L_{\mathrm{5\,GHz}}/L_{\mathrm{4400\,\AA}}$  radio-loudness index and reference number; Col.~5-6 -- $R_{\mathrm{2500\,\AA}} = L_{\mathrm{5\,GHz}}/L_{\mathrm{2500\,\AA}}$ radio-loudness index and reference number; Col.~7-8 -- L-band core brightness temperature (with uncertainty in parentheses) and reference number; Col.~9-10 -- L-band total flux density (with uncertainty in parentheses) and reference number; Col.~11-12 -- L-band core flux density (with uncertainty in parentheses) and reference number; Col.~13-14 -- L-band total radio power (with uncertainty in parentheses) and reference number; Col.~15-16 -- L-band core radio power (with uncertainty in parentheses) and reference number; Col.~17-18 -- estimated black hole mass and reference number. The literature references are the following. 0: This paper; 1:\citet{2023MNRAS.519.4047A}; 2: \citet{2015ApJ...804..118B}; 3: \citet{2021ApJ...909...80B}; 4: \citet{2022AA...659A.159S}; 5: \citet{2019AA...629A..68B}; 6: \citet{2020AA...635L...7B}; 7: \citet{2023AA...669A.134B}; 8: \citet{2024AA...684A..98C}; 9: \citet{2014AA...563A.111C}; 10: \cite{1998AJ....115.1693C}; 11: \citet{2011ApJ...739...56D}; 12: \citet{2003MNRAS.343L..20F}; 13: \citet{2008AA...484L..39F}; 14: \citet{2010AA...524A..83F}; 15: \citet{2015MNRAS.446.2921F}; 16: \citet{2015MNRAS.450L..57G}; 17: \citet{2018RNAAS...2..200G}; 18: \citet{2023PASA...40....4G}; 19: \citet{2015MNRAS.450L..34G}; 20: \citet{2021AA...647L..11I}; 21: \citet{2024MNRAS.528.5692K}; 22: \citet{2021ApJ...908..124L}; 23: \citet{2022ApJ...929...69L}; 24: \citet{2024A&A...685A.111L}; 25: \citet{2018ApJ...861...86M}; 26: \citet{2021AJ....161..207M}; 27: \citet{2012MNRAS.426L..91S}; 28: \citet{2021AA...655A..95S}; 29: \citet{2022AA...659A.159S}; 30: \citet{2019ApJ...873...35S}; 31: \citet{2020AA...643L..12S}; 32: \citet{2016ApJ...829...33Y}; 33: \citet{2022AA...662L...2Z}; 34: \citet{2013ApJ...773...44W}; 35: \citet{1997ApJ...475..479W}; 36: \citet{2015Natur.518..512W}; 37: \citet{2011A&A...531L...5F}; 38: \citet{2016MNRAS.463.3260C}  
    \label{tab:references}
\end{sidewaystable*}

\end{appendix}

\end{document}